\documentclass[%
 reprint,onecolumn,
 amsmath,amssymb,
 aps, showkeys
]{revtex4-2}
\usepackage{amsfonts}
\usepackage{graphicx}
\usepackage{dcolumn}
\usepackage{bm}
\usepackage{stmaryrd}
\usepackage{color}
\usepackage{cancel}
\usepackage{mathtools}
\usepackage{upgreek}
\graphicspath{{./figures/}}

\newcommand{\beq}{\begin{equation}}
\newcommand{\eeq}{\end{equation}}
\newcommand{\pfr}[2]{\ensuremath{\frac{\partial #1}{\partial #2}}}
\newcommand{\pfi}[2]{\ensuremath{{\partial #1}/{\partial #2}}}
\newcommand{\ep}{\varepsilon}

\newcommand{\mb}[1]{\mathbf{#1}}
\newcommand{\rr}{\mathfrak{r}}
\newcommand{\gr}{\mathrm{g}}
\newcommand{\mm}{\mathfrak{m}}
\newcommand{\ext}{\mathrm{ext}}

\newcommand{\mc}[1]{\mathcal{#1}}
\DeclareMathAlphabet\mathbfcal{OMS}{cmsy}{b}{n}
\newcommand{\oline}[1]{\overline{#1}}

\begin{document}

\preprint{APS/123-QED}

\title{Hydrodynamic theory of premixed flames under Darcy's law: Interfacial conditions and effects of nonunity Lewis number and heat loss}
\author{Prabakaran Rajamanickam, Joel Daou} 
 \email{joel.daou@manchester.ac.uk}
\affiliation{Department of Mathematics, University of Manchester, Manchester M13 9PL, United Kingdom}

\date{\today}

\begin{abstract}
Premixed flames propagating in porous media or Hele-Shaw channels are governed by Darcy's law, which accounts for the strong frictional forces imposed by the solid matrix or confining walls. Prior theoretical studies of such flames have typically employed phenomenological Markstein-type corrections and have assumed unity Lewis numbers and adiabatic conditions. In this work, we develop a rigorous hydrodynamic theory for premixed flames under Darcy's law that incorporates nonunity Lewis numbers and heat losses. Using large activation-energy asymptotics for a one-step Arrhenius chemistry model and a systematic multiple-scale analysis, we derive the interfacial jump conditions across the flame from first principles. The conventional continuity requirements of mass flux and pressure at an interface under Darcy's law acquire corrections to the finite thickness of the flame.  The adiabatic burning rate is shown to involve three distinct Markstein numbers, corresponding to curvature, tangential flow strain, and gravity-induced strain. The gravity term is unique to Darcy's law and has no counterpart in classical Navier–Stokes formulations. Moreover, the curvature Markstein number and the tangential strain Markstein number are found to be unequal, in contrast to the classical Navier--Stokes case where they coincide in the one-step  chemistry model. Explicit formulas for the Markstein numbers are provided, and the resulting new dispersion relation, linking the perturbation wave number $k$ to the growth rate $s$, takes the form $s = (a|k| - bk^2 - d|k|^3) / (1 + c|k|)$. This relation, applicable under Darcy's law, is to be  compared to the classical Clavin--Garcia dispersion relation derived from the Navier--Stokes equations. The theory provides a rigorous foundation for flame dynamics in strongly confined environments, with direct applications to porous media combustion and Hele-Shaw cell experiments. 
\end{abstract}


\maketitle


\section{Introduction}
\label{sec:intro}

Premixed flames propagating in permeable porous media exhibit fundamentally different behaviour from conventional flames, as the hydrodynamics in such systems are governed by Darcy’s law. Similarly, flames propagating in the narrow gap of a Hele--Shaw channel with nearly adiabatic walls can be qualitatively described by Darcy’s law, as demonstrated in several recent studies~\cite{fernandez2018analysis,martinez2019role,rajamanickam2024effect,daou2025hydrodynamic,chen2025three,DominguezGonzalez2024Pathway,DominguezGonzalez2024Numerical,DominguezGonzalez2022Stable}. Darcy’s law accounts for the substantial momentum losses caused by friction with the solid matrix or channel walls. As a result, physical parameters such as the permeability of the porous medium or the width of the Hele-Shaw channel play a central role in determining flame behaviour. The study of flame instability in Darcy-type flows was pioneered by Joulin and Sivashinsky~\cite{joulin1994influence}, who employed an Euler--Darcy model which combines heuristically the Eulers equation and the Darcy's law. This study has been revised in the recent works~\cite{miroshnichenko2020hydrodynamic,daou2025hydrodynamic} by accounting for a Markstein-type stabilisation at small wavenumbers, while also generalising the results to arbitrary magnitudes of an imposed flow. The three hydrodynamic instabilities, namely the Darrieus--Landau (DL), Saffman--Taylor (ST) and Rayleigh--Taylor (RT) instabilities, have  been delineated clearly in~\cite{daou2025hydrodynamic}. However, the Markstein-type stabilisation employed in~\cite{miroshnichenko2020hydrodynamic,daou2025hydrodynamic}, is phenomenological, rather than being based on an asymptotic analysis. In particular, these Markstein models lack a rigorous asymptotic derivation of 
the \textit{interfacial jump conditions} across the flame, as they neglect 
finite flame-thickness corrections to the continuity of mass flux and pressure. 
The present study addresses this issue by deriving asymptotically accurate 
interfacial jump conditions.

Furthermore, the so-called Markstein numbers, which characterise the flame responses to local curvature and flow straining, are typically regarded as prescribed parameters in stability analyses~\cite{miroshnichenko2020hydrodynamic,daou2025hydrodynamic}. A first-principles derivation of these non-dimensional numbers requires a rigorous multiple-scale analysis of a wrinkled premixed flame~\cite{clavin2016combustion}, a derivation first carried out by Clavin and Williams~\cite{clavin1982effects} using large activation-energy asymptotics for a one-step chemistry model governed by the Navier--Stokes equations. While correct interfacial jump conditions for conservation laws were derived systematically in~\cite{pelce1988influence,matalon1982flames} and a complete hydrodynamic theory established by Matalon and Matkowsky~\cite{matalon1982flames} and Clavin and Joulin~\cite{clavin1983premixed}, a key open question remains: do Markstein numbers derived under Navier–Stokes hydrodynamics remain applicable to flames governed by Darcy’s law?

In the absence of better alternatives, previous studies~\cite{al2019darrieus,radisson2022forcing}  have commonly applied Markstein-number expressions derived for unconfined flames even in strongly confined environments. This question was partially addressed in our preliminary study~\cite{rajamanickam2024hydrodynamic}, which investigated the internal flame structure for reactants with unity Lewis numbers while neglecting heat losses and interfacial jump conditions. That study highlighted that Markstein numbers and flame responses are modified at a fundamental level under Darcy’s law, differing significantly from those associated with unconfined flames. However, the unity Lewis-number and adiabatic approximations adopted in~\cite{rajamanickam2024hydrodynamic} are rather restrictive from a practical standpoint. Aside from methane–air and ethylene–air mixtures~\cite{rajamanickam2018two}, the unity Lewis-number assumption is rarely justifiable for arbitrary fuel mixtures. Likewise, the adiabatic approximation seldom holds in practice, particularly in Hele–Shaw configurations, where conductive heat losses to the channel walls are usually very severe~\cite{martinez2019role,han2021effect}. One of the main objectives of the present work is thus to derive Markstein-number formulas that relax both of these assumptions, thereby providing results of broader relevance to both theoretical and experimental studies.

The remainder of this paper is organised as follows. Section~\ref{sec:form}
presents the governing equations for curved premixed flames under Darcy’s law 
and outlines the underlying physical approximations. Section~\ref{sec:asymptotics} then details the multiple-scale analysis used 
to derive the hydrodynamic model, explicit Markstein numbers, and jump conditions. 
These results generalize the model established in~\cite{rajamanickam2024hydrodynamic} 
by providing a more accurate and general description.  Our methodological approach parallels classical multiple-scale asymptotic 
studies for conventional Navier--Stokes flames, which were subsequently extended 
to account for non-unity Lewis numbers~\cite{clavin1982effects,pelce1988influence,matalon1982flames,clavin1983premixed,clavin1983influence} 
and heat losses~\cite{clavin1985effect,keller1994transient,matalon2009multi}. 
As the underlying algebraic derivations are quite lengthy, readers interested 
only in the final results may skip this section. Section~\ref{sec:summary} 
then presents the summary of the hydrodynamic model. Finally, Section~\ref{sec:stability} 
presents the linear stability analysis of premixed flames under an imposed 
flow and gravity, yielding a new dispersion relation that generalises and 
refines previous formulations~\cite{daou2025hydrodynamic,rajamanickam2026flame} 
by incorporating the correct interfacial jump conditions.

\section{Problem formulation}
\label{sec:form}

Consider a premixed flame propagating through a fuel-lean reacting gas mixture.  The fresh (unburnt) mixture is assumed to have constant values of  density $\rho_u$, viscosity $\mu_u$, permeability $\kappa_u$ and thermal diffusivity $D_{T,u}$.  Similarly, the burnt gas mixture immediately behind the flame is characterised by constant properties $\rho_b$, $\mu_b$, $\kappa_b$ and $D_{T,b}$. In the case of Hele-Shaw channels, the permeabilities of the unburnt and burnt gases are in fact identical, given by $\kappa_u=\kappa_b=H^2/12$, where $H$ denotes the channel width. The reaction rate per unit volume is assumed to follow an Arrhenius law of the form $\rho B Y_F e^{-E/RT}$, where $B$ is the pre-exponential factor and $E/R$ is the activation energy. The adiabatic flame temperature is defined as $T_{ad}=T_u(1+q)$, where $q$ quantifies the heat released by the chemical reaction. The Lewis number of the fuel, $Le$, equal to the ratio of thermal diffusivity to fuel diffusion coefficient,  is assumed to be constant. Conductive heat losses are modelled by the presence of a  heat-loss rate (per unit volume) term which takes the form $\rho_u c_p K (T-T_u)$ as in Newton's law of cooling, with $c_p$ being the specific heat at constant pressure and $K$ a heat-transfer coefficient.

The characteristic length scale $L$ associated with flame wrinkling is assumed to be much larger than the flame thickness $\delta_L \equiv D_{T,u}/S_L$. Here, $S_L$ represents the planar flame speed in the large activation-energy 
asymptotic limit, given by the expression $S_L^2 = 2\beta^{-2} B D_{T,b} Le (\rho_b^2/\rho_u^2) e^{-E/RT_{ad}}$, 
which is valid in the limit $\beta \to \infty$. The parameter $\beta = E(T_{ad}-T_u)/RT_{ad}^2$ 
denotes the Zeldovich number. Our analysis is based on expansions in a small  parameter $\ep$ defined as
\begin{equation}
    \ep = \frac{\delta_L}{L} \ll 1 \,. \nonumber
\end{equation}
For convenience, physical quantities are non-dimensionalised using $L$ as the characteristic length scale, $L/S_L$ as the time scale, $S_L$ as the velocity scale, and $\mu_u D_{T,u}/(\ep\kappa_u)$ as the hydrodynamic pressure scale. All fluid properties, including fluid density and transport properties, are non-dimensionalised using their respective values in the unburnt gas. The scaled fuel mass fraction  is defined by $y_F=Y_F/Y_{F,u}$, and the non-dimensional temperature   by $\theta=(T-T_u)/(T_{ad}-T_u)$, such that $y_F, \theta \in[0,1]$. The main objective of this paper is to describe flame propagation in the large-activation-energy asymptotic limit $\beta\to\infty$. The analysis will thus focus on the near equi-diffusional and near-adiabatic regimes, characterised by the conditions
\begin{align}
    l\equiv \beta(Le-1)\sim O(1) \,, \qquad  \kappa \equiv  \beta K\delta_L/S_L\sim O(1),   \nonumber
\end{align}
where $l$ is the reduced Lewis number and $\kappa$ is the heat-loss parameter (not to be confused with the permeabilities $\kappa_u$ and $\kappa_b$ introduced above and which will not appear in the non-dimensional problem).  Furthermore, instead of working directly with the fuel mass fraction $y_F$,  we shall work with the enthalpy variable
\begin{equation}
    h \equiv\beta(\theta+y_F-1)    \nonumber
\end{equation}
as it is common in near equi-diffusional analyses of flames~\cite[p.~38]{buckmaster1983lectures}.

 \begin{figure}
\centering
\includegraphics[scale=0.57]{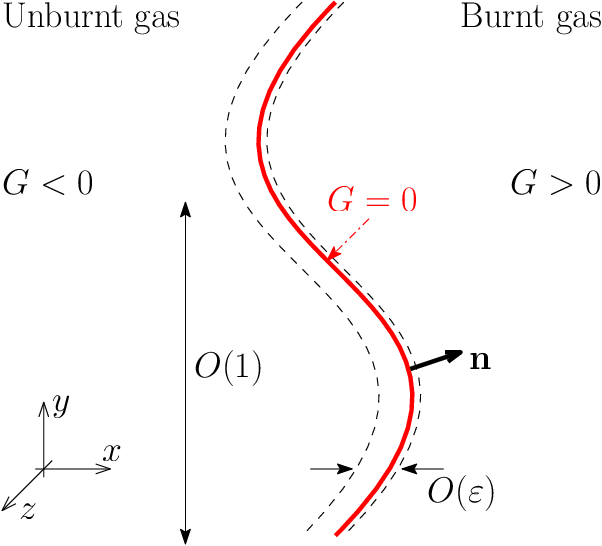}
\caption{Schematic illustration of a curved premixed flame of thickness $O(\ep)$ propagating into a fresh mixture. The level set $G=0$ of the field $G(\mb x,t)$ represents both the flame-front location as perceived from the outer hydrodynamic region and the location of the reaction sheet.} 
\label{fig:sch}
\end{figure}
We introduce the well-known $G$-equation~\citep{williams1985turbulent},
\begin{equation}
   \rho\left(\pfr{G}{t}+\mb v\cdot \nabla G\right) = \dot m |\nabla G|  \label{Geq}
\end{equation}
where $\dot m(\mb x,t)$ denotes the non-dimensional local burning rate (scaled by $\rho_u S_L$), i.e., the normal mass flux crossing a given level set $G(\mb x,t)=\text{const}$. For any such level set of the function $G(\mb x,t)$,   the local unit normal vector pointing towards the burnt gas is given by  $\mb n=\nabla G/|\nabla G|$. In the present work, we identify the level set $G=0$ as representing both the flame-front location from the perspective of the outer hydrodynamic region and the location of the reaction sheet. Alternative formulations in which these two locations are associated with distinct level sets can be found in~\cite{bechtold2001dependence,clavin2016combustion,giannakopoulos2015consistent,giannakopoulos2019consumption,rajamanickam2024hydrodynamic}. A schematic illustration of the premixed flame  involving 
the pertinent scales of the non-dimensional problem is shown in Fig.~\ref{fig:sch}. In the regions on either side of the reaction sheet ($G < 0$ and $G > 0$), the governing equations are assumed to take the form
\begin{align}
    \pfr{\rho}{t} + \nabla \cdot(\rho \mb v)& = 0 \,,  \label{conteq}\\
    \mu \mb v = -\nabla p &+ \rho \mb g \,,  \label{darcy}\\    
    \rho \pfr{\theta}{t} + \rho \mb v\cdot \nabla \theta & =   \ep\nabla\cdot(\lambda \nabla \theta) \,, \label{theta}\\ 
    \rho \pfr{h}{t} + \rho \mb v\cdot \nabla h & =  \ep\nabla\cdot[\lambda \nabla (h+l\theta)] -\frac{\kappa\theta}{\ep} \,, \label{theta}\\  
    \rho = \rho(\theta), \quad \mu &=\mu (\theta), \quad \lambda = \lambda(\theta) 
\end{align}
where $\mb g$ is the non-dimensional gravity vector, whose magnitude, $|\mb g|=\rho_u g \kappa_u/\mu_u S_L$, measures the strength of the gravitational force.  The non-dimensional resistance function $\mu=\mu(\theta)$ accounts for variations in both fluid viscosity and permeability, while $\lambda(\theta)$ denotes the non-dimensional thermal conductivity. With the exception of the temperature equation, all of the above equations are in fact valid everywhere, including inside the thin diffusive–reactive zone of thickness $O(\ep/\beta)$.  The above governing equations, including Darcy's law, have been widely employed in recent studies to describe flame dynamics within Hele-Shaw cells~\cite{fernandez2018analysis,martinez2019role,rajamanickam2024effect,daou2025hydrodynamic,chen2025three,DominguezGonzalez2024Pathway,DominguezGonzalez2024Numerical,DominguezGonzalez2022Stable}. This depth-averaged system is formally applicable under the scale separation assumption $H \sim \delta_L \ll L$, where cross-gap variations remain negligible.

Across the reaction sheet (not the flame front), the physical variables satisfy a set of well-established jump conditions~\cite[p.~39]{buckmaster1983lectures}. These are most conveniently expressed in terms of a stretched inner coordinate $\zeta$, defined by
\begin{equation} 
    \zeta  =  \frac{G}{\ep} \,.   \label{smallcoord}
\end{equation}
Specifically, at the reaction sheet, corresponding to $\zeta=0$, the jump conditions read  
\begin{align}
  \{\!\{ \mb v \}\!\}=\{\!\{ p \}\!\}=\{\!\{ \theta \}\!\}=\{\!\{ h \}\!\}= 0 \,, \quad  
  \lambda|\nabla G|\left\{\!\left\{\pfr{\theta}{\zeta}\right \}\!\right\}+e^{h/2}=  0 \,, \quad \left\{\!\left\{\pfr{h}{\zeta}\right \}\!\right\}+l\left\{\!\left\{\pfr{\theta}{\zeta}\right \}\!\right\}=0  \label{sheetjump}
\end{align}
where the (reaction-sheet) jump operator is defined as $\{\!\{\varphi  \}\!\}\equiv \varphi |_{\zeta=0^+}-\varphi |_{\zeta=0^-}$. Finally, as $\zeta\to\pm\infty$, the inner solutions are required to match smoothly with the corresponding outer solutions. For later convenience, we introduce the following bracket notation to denote the jump of any property $\varphi$ across the entire flame thickness:
\begin{equation}
    \llbracket \varphi \rrbracket = \varphi(+\infty) - \varphi(-\infty), \nonumber
\end{equation}
where, for example, $\llbracket \theta \rrbracket = 1 + O(1/\beta)$. Indeed, a primary objective of this work is to determine these interfacial jumps for various physical quantities, allowing the flame to be treated on the hydrodynamic $L$-scale as a hydrodynamic discontinuity.

\section{Multiple-scale analysis}\label{sec:asymptotics}

In this section, a multiple-scale analysis is carried out to describe the 
asymptotic structure of a wrinkled premixed flame in the limit $\epsilon\to 0$. 
Before proceeding, readers who are not interested in the rather lengthy algebraic 
derivations may skip directly to Section~\ref{sec:summary}, where a summary of 
the main results is provided for convenience. 

To pursue this derivation,  we follow the approach of Keller and 
Peters~\cite{keller1994transient}, whose methodology carries the distinct 
advantage that the analysis is performed in a fixed Cartesian coordinate 
system $(\mb x,t)$ rather than a flame-attached frame. Within this framework, all physical variables, except for the level-set function $G$ (and hence the normal vector $\mb n$), are assumed to depend on both the large-scale hydrodynamic variables $(\mb x,t)$ and the small-scale flame coordinate $\zeta$. Specifically, for any physical variable $\varphi$, we introduce the  expansion
\begin{equation}
    \varphi = \varphi_0(\zeta,\mb x,t) + \ep \varphi_1(\zeta,\mb x,t) + \cdots . \nonumber
\end{equation}
The level-set function $G=G(\mb x,t)$, which by definition cannot depend on $\zeta$, is assumed to be known to arbitrary accuracy in powers of $\ep$ and is therefore not expanded, following~\cite{clavin2011curved,clavin2016combustion,rajamanickam2024hydrodynamic}. Within the multiple-scale framework, derivatives with respect to the large-scale coordinates $(x^i,t)$ transform according to
\begin{equation}
    \pfr{}{x^i}\mapsto \frac{1}{\ep} \pfr{G}{x^i}\pfr{}{\zeta} + \pfr{}{x^i} \,, \qquad  \pfr{}{t}\mapsto \frac{1}{\ep} \pfr{G}{t}\pfr{}{\zeta} + \pfr{}{t} \,.   \nonumber
\end{equation}
Applying these transformations to the continuity equation~\eqref{conteq}, and combining the result with the $G$-equation~\eqref{Geq}, yields the modified continuity relation
\begin{equation} \label{modcont}
 \pfr{\rho}{t}+\nabla\cdot(\rho\mb v) = -\frac{|\nabla G|}{\ep} \pfr{\dot m}{\zeta} \,.
\end{equation}
Physical variables   found to be independent of the inner coordinate $\zeta$,  can  be identified as   \textit{outer variables} associated with the hydrodynamic region.

\subsection{Structure of the locally planar flame}

\begin{figure}
\centering
\includegraphics[scale=0.5]{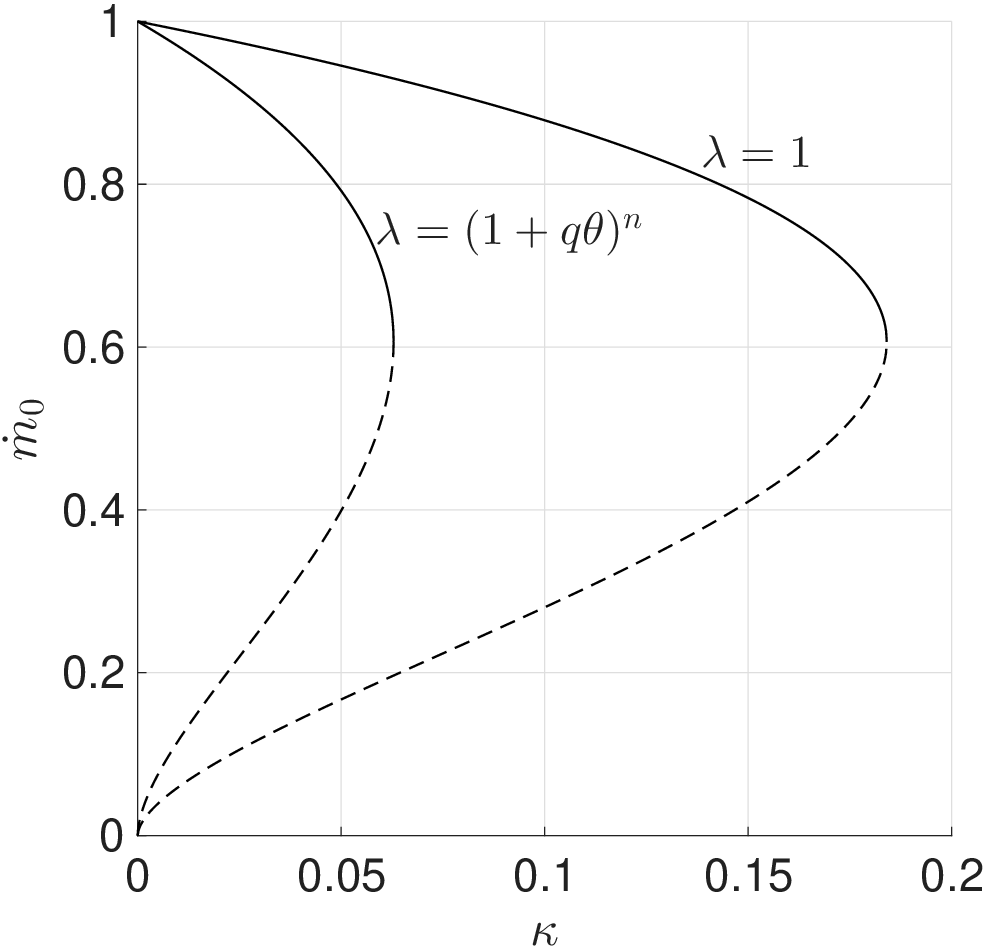}
\caption{Leading-order burning rate $\dot m_0 = \dot M_0$ as a function of the heat-loss parameter $\kappa$. Accounting for variable thermal conductivity (with $q = 5$ and $n = 0.7$) reduces the extinction limit. The solid lines correspond to the stable branches and approach the adiabatic speed $\dot m_0 = 1$ as $\kappa \to 0$.} 
\label{fig:burningrate}
\end{figure}

At leading order, the governing equations reduce to
\begin{align} \label{leading}
    \pfr{\dot m_0}{\zeta} =    \pfr{p_0}{\zeta} =0 \,, \quad 
    \frac{\dot m_0}{|\nabla G|} \pfr{\theta_0}{\zeta} = \pfr{}{\zeta}\left(\lambda_0\pfr{\theta_0}{\zeta}\right) \,, \quad \frac{\dot m_0}{|\nabla G|} \pfr{h_0}{\zeta} = \pfr{}{\zeta}\left(\lambda_0\pfr{(h_0+l\theta_0)}{\zeta}\right)-\frac{\kappa\theta_0}{|\nabla G|^2} \,.
\end{align}
The solution of this system, subject to the jump conditions~\eqref{sheetjump}, is given by
\begin{align} 
\label{leadingsol}
p_0 &= P_0(\mb x,t) \,, \quad 
\dot m_0 = \dot M_0 = \text{const.} \,, \quad
\theta_0 = 
\begin{cases}
\displaystyle e^{\dot m_0\hat\zeta/|\nabla G|} \,, & \zeta < 0  \\[8pt]
1 \,, & \zeta > 0 
\end{cases} 
\\[12pt]
h_0 &=-l\theta_0\ln\theta_0 - \frac{\kappa}{\dot{m}_0^2} 
\begin{cases}
\displaystyle \theta_0\int_{\theta_0}^{1}\frac{\lambda_0'}{\theta_0'}\,d\theta_0' + \lambda_f\theta_0 + \int_{0}^{\theta_0}\lambda_0'\,d\theta_0' \,, & \zeta < 0  \\[12pt]
\displaystyle \lambda_f + \int_{0}^{1}\lambda_0\,d\theta_0 + \frac{\dot{m}_0 \zeta}{|\nabla G|} \,, & \zeta > 0  
\end{cases}
\end{align}
where $\hat\zeta = \int_0^\zeta d\zeta/\lambda_0$  and $\lambda_f \equiv \lambda_0(1)$. We introduce the primed notation to denote evaluation at the dummy variable 
$\theta_0'$, such that $\lambda_0' \equiv \lambda_0(\theta_0')$ and, more 
generally, $\varphi' \equiv \varphi(\theta_0')$ hereafter. The 
leading-order outer burning rate $\dot M_0$ and outer pressure $P_0(\mb x,t)$  remain constant across the flame thickness. Specifically, the leading-order burning rate $\dot m_0=\dot M_0$ is constant and is determined from
\begin{equation}   \label{eq:m0exp}
    \dot m_0^2 \ln \dot m_0^2 = -\frac{\kappa}{e\kappa_\ext}  \qquad \text{with} \qquad  \kappa_\ext\equiv  \frac{1}{e\left(\lambda_f+\int_0^1\lambda_0 d\theta_0 \right)} \,.
\end{equation}
The solution $\dot m_0$ to this equation exists and is double-valued   in the range $0\le\kappa<\kappa_\ext$, with the larger burning-rate solution being the physically relevant, stable one.  Along the stable upper branch, the burning rate approaches the adiabatic value, $\dot m_0 \to 1$, in the limit $\kappa \to 0$. For a constant thermal conductivity ($\lambda_0 = 1$), the extinction limit predicted in~\eqref{eq:m0exp} reduces to
\begin{equation}
    \kappa_{\text{ext}} = \frac{1}{2e} \,,  \nonumber
\end{equation}
whereas for a temperature-dependent thermal conductivity $\lambda_0 = (1+q\theta_0)^n$, it becomes
\begin{equation}   
    \kappa_{\text{ext}} = \frac{q(n+1)}{e \left[ (1+q)^n (1+q(n+2)) - 1 \right]} \,.   \nonumber
\end{equation}
In particular, for the typical values $q = 5$ and $n = 0.7$, a plot of
$\dot m_0$ versus $\kappa$ is provided in Fig.~\ref{fig:burningrate}. The figure shows that accounting for the temperature dependence of thermal conductivity substantially reduces the extinction limit, by roughly a factor of three in this case. For later reference, we introduce the quantity $h_f$, which represents the $O(1/\beta)$ reduction in the flame temperature from its adiabatic value due to heat loss:
\begin{equation}
    1 - \theta_f = \frac{h_f}{\beta} \,, \qquad h_f \equiv h_0(0,\mb x,t) = \ln \dot m_0^2 = -\frac{\kappa}{\dot m_0^2}\left(\lambda_f + \int_0^1\lambda_0 \, d\theta_0\right) .
\end{equation}
The two terms in the final expression for $h_f$ account for the integrated heat loss in the burned and unburned gas regions, respectively~\cite{matalon2009multi}. 

\subsection{Leading-order flow field}\label{sec:leadingorderflow}

In the previous subsection, we found that the leading-order pressure field is $p_0=P_0(\mb x,t)$ everywhere, which  implies that it has zero jump   across the flame. The corresponding velocity field $\mb v_0(\zeta,\mb x,t)$ can be determined by examining the momentum equation at the next order, which is given by
\begin{equation}
    -\mu_0 \mb v_0 = \nabla P_0 - \rho_0 \mb g + \nabla G \pfr{p_1}{\zeta} \,. \label{leadingmomentum}
\end{equation}
The pressure gradient term $\pfi{p_1}{\zeta}$ can be eliminated by taking the   cross product of~\eqref{leadingmomentum} with the unit normal $\mb n$, which gives $-\mu_0 \mb v_0 \times \mb n = \nabla P_0 \times \mb n - \rho_0 \mb g \times \mb n$. Taking a second cross product with $\mb n$ from the left isolates the tangential component of $\mb v_0$ yielding $\mathbfcal P\mb v_0=\mb n\times (\mb v_0\times\mb n)$, where $\mathbfcal P =\mb I -\mb n\otimes\mb n$ denotes the projection operator onto the local tangent plane of the flame surface. The normal component $\mb v_0\cdot\mb n$ is obtained from the leading-order $G$-equation $\rho_0 (\pfi{G}{t}+\mb v_0\cdot \nabla G) =  \dot m_0|\nabla G|$. Combining the normal and tangential components yields
\begin{equation} \label{v0Eq} 
    \mb v_0 = \left(\frac{\dot m_0}{\rho_0} - \frac{1}{|\nabla G|}\pfr{G}{t}\right) \mb n - \frac{\mathbfcal P}{\mu_0}(\nabla P_0 - \rho_0 \mb g) \,.
\end{equation}

It is convenient to introduce an auxiliary outer velocity fields $\mb V_0^\pm(\mb x,t)$,  which satisfy
\begin{align}
    \nabla \cdot \mb V_0^- =0, &\qquad -\mb V_0^- = \nabla P_0 - \mb g \qquad \text{for} \quad G<0, \label{outer0unburnt}\\
    \nabla \cdot \mb V_0^+ =0, &\qquad -\mu_f\mb V_0^+ = \nabla P_0 -\rho_f \mb g\qquad \text{for} \quad G>0, \label{outer0burnt}
\end{align}
along with the kinematic constraints
\begin{equation}
    \frac{\tilde D^-G}{\tilde D t}\equiv \pfr{G}{t}+\mb V_0^- \cdot \nabla G =  \dot M_0|\nabla G|, \qquad  \frac{\tilde D^+G}{\tilde D t}\equiv  \pfr{G}{t}+\mb V_0^+ \cdot \nabla G =  \frac{\dot M_0}{\rho_f}|\nabla G|.  \label{outer0G}
\end{equation}
Proceeding as in the derivation of~\eqref{v0Eq}, the last three equations in~\eqref{outer0unburnt}–\eqref{outer0G} yield
\begin{equation} \label{V0Eq}
    \mb V_0^- = \left(\dot M_0 - \frac{1}{|\nabla G|}\pfr{G}{t}\right) \mb n - \mathbfcal P(\nabla P_0 -  \mb g), \quad \mb V_0^+ = \left(\frac{\dot M_0}{\rho_f} - \frac{1}{|\nabla G|}\pfr{G}{t}\right) \mb n - \frac{\mathbfcal P}{\mu_f}(\nabla P_0 -  \rho_f \mb g).
\end{equation}
Combining~\eqref{v0Eq} and~\eqref{V0Eq} then gives 
\begin{equation} 
    \mb v_0 = \begin{cases}
        \displaystyle \mb V_0^- + \frac{1-\rho_0}{\rho_0}  \dot m_0\mb n - \frac{\mu_0-1}{\mu_0}\mathbfcal P\mb V_0^- - \frac{1-\rho_0}{\mu_0}\mathbfcal P\mb g \quad \qquad &\text{as} \quad \zeta <0,\\ 
        \displaystyle \mb V_0^+ \quad \qquad &\text{as} \quad \zeta >0.
    \end{cases}  \label{innerflow}
\end{equation}
This relation shows that $\mb v_0-\mb V_0^-$ decays exponentially as $\zeta\to-\infty$, due to the dependence of $\rho_0$ and $\mu_0$ on $\theta_0$ which decays exponentially in this limit. As for the quantity $\mb v_0-\mb V_0^+$, it vanishes identically for $\zeta>0$. Consequently, the flow field $(\mb V_0^\pm,P_0)$, which is incompressible and satisfies Darcy’s law for $G\neq0$ according to~\eqref{outer0unburnt}–\eqref{outer0burnt},  represents indeed the leading-order outer flow. The continuity of $\mb v_0$ at $\zeta=0$ yields
\begin{equation}
   \mb V_0^+ = \mb V_0^- + \frac{1-\rho_f}{\rho_f}\dot m_0 \mb n - \frac{\mu_f-1}{\mu_f}\mathbfcal P\mb V_0^- - \frac{1-\rho_f}{\mu_f}\mathbfcal P\mb g.  \label{V0plusminus}
\end{equation}
Thus, the leading-order outer flow field satisfies the following jump conditions across the flame,
\begin{equation}
    \llbracket P_0 \rrbracket =0, \qquad \llbracket \mb V_0\cdot \mb n \rrbracket = \frac{1-\rho_f}{\rho_f}\dot m_0, \qquad  \llbracket \mu \mathbfcal P \mb V_0 \rrbracket = \llbracket \rho \mathbfcal P \mb g \rrbracket. \label{leadingorderjump}
\end{equation}
The last expression implies that the tangential component $\mathbfcal P\mb V_0$ of the outer velocity field is discontinuous across the flame front either due to viscosity jumps $\llbracket \mu  \rrbracket \neq 0$ or to jumps in the gravitation term $\llbracket \rho \mathbfcal P \mb g  \rrbracket \neq \mb 0$. The leading-order composite expansion~\eqref{innerflow} describes how these discontinuities are smoothed within the inner flame zone. Furthermore, expression~\eqref{V0plusminus} allows us to relate the two operators $\tilde D^\pm/\tilde D t$ through the relation
\begin{equation}
    \frac{\tilde D^+}{\tilde Dt} = \frac{\tilde D^-}{\tilde Dt} + \left(\frac{1-\rho_f}{\rho_f}\dot m_0 \mb n - \frac{\mu_f-1}{\mu_f}\mathbfcal P\mb V_0^- - \frac{1-\rho_f}{\mu_f}\mathbfcal P\mb g \right)\cdot \nabla \,\,. \label{Dplusminus}
\end{equation}

\subsection{First correction to the burning rate}

The continuity equation~\eqref{modcont} at the first order implies
\begin{align}
    |\nabla G|\pfr{\dot m_1}{\zeta} = - \pfr{\rho_0}{t} - \nabla\cdot(\rho_0\mb v_0) = \begin{cases}
        \displaystyle\frac{\tilde D^- }{\tilde D t}(1-\rho_0)  - \nabla \cdot [\rho_0(\mb v_0 -\mb V_0^-)], \quad &\zeta<0,\\
        \displaystyle 0, \quad &\zeta<0,
    \end{cases} \label{m1equation}
\end{align}
where the second equality for $\zeta<0$ is obtained by adding and subtracting $\nabla\cdot(\rho_0\mb V_0^-)$ to the middle expression, using $\nabla\cdot\mb V_0^-=0$, and  the notation introduced in~\eqref{outer0G}. The solution for $\dot m_1$ in the burnt-gas region $(\zeta>0)$ is straightforward, i.e., $\dot m_1 = \dot M_1^+(\mb x,t)$. For $\zeta<0$, we integrate~\eqref{m1equation} from $\zeta=-\infty$ to an arbitrary location $\zeta$. Since the differential operators on the right-hand side involve  only outer variables, the integration can be commuted with these  operators. Furthermore, the integration can be conveniently carried out by changing variables from $\zeta$ to $\theta_0=e^{\dot m_0\hat\zeta/|\nabla G|}$, as given by~\eqref{leadingsol}, so that $d\zeta = d\theta_0\, |\nabla G| \lambda_0/\theta_0\dot m_0 $. Performing this integration after using the expression~\eqref{innerflow} for $\mb v_0-\mb V_0^-$, we obtain  
\begin{align}
    \dot m_1(\zeta,\mb x,t) = \dot M_1^-(\mb x,t) + \frac{1}{\dot m_0|\nabla G|} \left\{\frac{\tilde D^- }{\tilde D t}(\hat{\mc I_1} |\nabla G|)  - \dot m_0\nabla \cdot(\hat{\mc I_1}\nabla G) + \nabla\cdot \left[|\nabla G|\mathbfcal P (\hat{\mc I_2} \mb V_0^-+\hat{\mc I_3} \mb g) \right]\right\},  \quad \zeta<0,\label{m1}
\end{align}
where $\dot M_1^-(\mb x,t)$ is an integration constant, and the functions $\hat{\mc I}_1$, $\hat{\mc I}_2$, and $\hat{\mc I}_3$ are defined by
\begin{align} 
    \hat{\mc I_1} (\theta_0)=  \int_0^{\theta_0} \frac{\lambda_0'}{\theta_0'}(1-\rho_0') d\theta_0' \,, \quad  \nonumber \hat{\mc I_2}(\theta_0) = \int_0^{\theta_0} \frac{\rho_0'\lambda_0'}{\mu_0'\theta_0'}(\mu_0'-1)d\theta_0' \,,\quad \nonumber
   \hat{\mc I_3}(\theta_0) = \int_0^{\theta_0} \frac{\rho_0'\lambda_0'}{\mu_0'\theta_0'}(1-\rho_0)d\theta_0' \,.
\end{align}
The function $\dot m_1-\dot M_1^-$ decays exponentially as $\zeta\to-\infty$ since $\theta_0\to0$ in the above integrals, and the function $\dot m_1-\dot M_1^+$ is  identically zero for $\zeta>0$. Consequently, $\dot M_1^\pm(\mb x,t)$ represent the first-order corrections to the burning rate in the outer hydrodynamic region. Evaluated at the location of the reaction sheet, $\zeta=0$  where $\theta_0 = 1$, the integrals $(\hat{\mc I_1},\hat{\mc I_2}, \hat{\mc I_3})$ become $(\mc I_1,\mc I_2,\mc I_3) \equiv (\hat{\mc I_1}(1),\hat{\mc I_2}(1), \hat{\mc I_3}(1))$ and come out of the derivatives in~\eqref{m1}. We thus obtain for $\dot m_1$ evaluated at $\zeta=0$ the expression
\begin{align}
   \dot m_1(0,\mb x,t)  = \dot M_1^-(\mb x,t) + \frac{1}{\dot m_0|\nabla G|}\left\{\mc I_1 \frac{\tilde D^-|\nabla G|}{\tilde D t} -\dot m_0\mc I_1 \nabla^2 G + \mc I_2\nabla\cdot \left(|\nabla G|\mathbfcal P \mb V_0^- \right)  +\mc I_3\nabla\cdot \left(|\nabla G|\mathbfcal P \mb g \right)\right\}.  \label{m1 equation}
\end{align}
The continuity of $\dot m_1$ at the reaction sheet $\zeta=0$ then reveals
\begin{equation}
   \dot M_1^+ = \dot M_1^- + \frac{1}{\dot m_0|\nabla G|}\left\{\mc I_1 \frac{\tilde D^-|\nabla G|}{\tilde D t} -\dot m_0\mc I_1 \nabla^2 G + \mc I_2\nabla\cdot \left(|\nabla G|\mathbfcal P \mb V_0^- \right)  +\mc I_3\nabla\cdot \left(|\nabla G|\mathbfcal P \mb g \right)\right\},  \label{M1plusminus}
\end{equation}
which is an interesting result since the jump in the burning rate, $\llbracket \dot M_1 \rrbracket$ across the flame now contains additional Darcy-specific contributions, associated with the terms involving $\mc I_2$ and $\mc I_3$.

The remaining objective of this subsection is to determine the function $\dot M_1^-$. In order to do that, we need to examine the first-order temperature and enthalpy equations together:
\begin{align}
& |\nabla G|^2 \pfr{}{\zeta} \left( \lambda_0 \pfr{\theta_1}{\zeta} + \lambda_1 \pfr{\theta_0}{\zeta} \right) 
  - \dot m_0 |\nabla G| \pfr{\theta_1}{\zeta} 
= |\nabla G| \pfr{(\dot m_1 \theta_0)}{\zeta} 
  + \pfr{(\rho_0\theta_0)}{t} 
  + \nabla \cdot (\rho_0 \mb{v}_0 \theta_0) \nonumber \\
&\qquad  {} - \nabla \cdot \left( \nabla G \, \lambda_0 \pfr{\theta_0}{\zeta} \right) 
  - \nabla G \cdot \pfr{}{\zeta} \left( \lambda_0 \nabla\theta_0 \right) \,, 
  \label{temp1} \\[4pt]
& |\nabla G|^2 \pfr{}{\zeta} \left( \lambda_0 \pfr{(h_1+l\theta_1)}{\zeta} 
  + \lambda_1 \pfr{(h_0+l\theta_0)}{\zeta} \right) 
  - \dot m_0 |\nabla G| \pfr{h_1}{\zeta} - \kappa\theta_1 
= |\nabla G| \pfr{(\dot m_1 h_0)}{\zeta} 
  + \pfr{(\rho_0h_0)}{t} 
  + \nabla \cdot (\rho_0 \mb{v}_0 h_0) \nonumber \\
&\qquad  {} - \nabla \cdot \left( \nabla G \, \lambda_0 \pfr{(h_0+l\theta_0)}{\zeta} \right) 
  - \nabla G \cdot \pfr{}{\zeta} \bigl[ \lambda_0 \nabla(h_0+l\theta_0) \bigr] \,. 
  \label{enthal1}
\end{align}
To facilitate the calculations below, we rewrite the right-hand sides of equation~\eqref{temp1} and~\eqref{enthal1}, applicable in the preheat zone $\zeta<0$, as
\begin{align}
    RHS_{\theta_1}^- &= |\nabla G| \pfr{(\dot m_1 \theta_0)}{\zeta} 
  + \frac{\tilde D^-(\rho_0\theta_0)}{\tilde D t} - \dot m_0\nabla\cdot\left(\rho_0\theta_0\mb n\right)
  -\nabla\cdot\left\{\frac{\rho_0\theta_0}{\mu_0}\left[(\mu_0-1)\mathbfcal P\mb V_0^- + (1-\rho_0)\mathbfcal P\mb g\right]\right\} \nonumber \\ & \quad + \pfr{}{\zeta}(\lambda_0\theta_0\ln\theta_0)\mb n\cdot\nabla|\nabla G|, \label{RHStheta1}\\
 RHS_{h_1}^- & = |\nabla G| \pfr{(\dot m_1 h_0)}{\zeta} 
   + \frac{\tilde D^-(\rho_0h_0)}{\tilde D t}
 -\dot m_0\nabla\cdot(\rho_0h_0\mb n)- \frac{\kappa}{\dot m_0}\nabla\cdot\left(\mb n\int_0^{\theta_0}\lambda_0' d\theta_0'\right) \nonumber \\
&\quad {} -\nabla\cdot\left\{\frac{\rho_0h_0}{\mu_0}\left[(\mu_0-1)\mathbfcal P\mb V_0^- + (1-\rho_0)\mathbfcal P\mb g\right]\right\} + \pfr{}{\zeta}\left[\lambda_0\ln\theta_0\left(h_0 +\frac{\kappa}{\dot m_0^2}\int_0^{\theta_0}\lambda_0'd\theta_0'\right)\right]\mb n\cdot\nabla|\nabla G|,   \label{RHSh1}
\end{align}
where we used the relations $\lambda_0\pfi{\theta_0}{\zeta}=\dot m_0\theta_0/|\nabla G|$ and $\nabla\theta_0 = -\theta_0\ln\theta_0\nabla(|\nabla G|)/|\nabla G|$ in the first equation and $\lambda_0\pfi{(h_0+l\theta_0)}{\zeta}=[h_0 +(\kappa/\dot m_0^2)\int_0^{\theta_0}\lambda_0'd\theta_0')]\dot m_0/|\nabla G|$ and $\nabla(h_0+l\theta_0)=-[h_0 +(\kappa/\dot m_0^2)\int_0^{\theta_0}\lambda_0'd\theta_0')]\ln\theta_0 \nabla(|\nabla G|)/|\nabla G|$ in the second. 

The solution for $\theta_1$ is formally written as 
\begin{align}
    \theta_1 = 
    \begin{cases}
        \displaystyle 
        \frac{\theta_0}{\lambda_0|\nabla G|^2} 
        \int_0^{\zeta}
        \left[ \frac{1}{\theta_0} \right]_{\zeta'}
        \int_{-\infty}^{\zeta'}
        \bigl[  \, RHS_{\theta_1}^- \bigr]_{\zeta''}
        \, d{\zeta}'' \, d{\zeta}' \,, 
        & \zeta < 0, \\[4ex]
        \displaystyle 
        0 \,, 
        & \zeta > 0,
    \end{cases}  \label{theta1sol}
\end{align}
where we used the relation $\lambda_1=\theta_1d\lambda_0/d\theta_0$ in carrying out the integration. This solution tends to zero as $\zeta \to \pm \infty$, thereby satisfying the 
matching conditions with the outer solution. It also satisfies the continuity  requirement $\llbracket \theta_1 \rrbracket = 0$ at the reaction sheet. Additional 
requirements on $\theta_1$ stem from the two derivative jump conditions 
in~\eqref{sheetjump}; these are yet to be enforced, as they depend on $h_1$. 
These two conditions will determine the remaining unknowns, namely, 
$\dot M_1^-(\mb x,t)$ and $h_1(0,\mb x,t)$.

The explicit solution for $h_1$ is needed only in the burnt gas $(\zeta>0)$ where  $\theta_0=1$, $\lambda_0=\lambda_f$, $\rho_0=\rho_f$, $\theta_1=0$, $\lambda_1=0$  and $h_0=h_f -\kappa\zeta/\dot m_0|\nabla G|$. In the burnt gas, equation~\eqref{enthal1} simplifies to
\begin{align}
    \lambda_f|\nabla G|^2 \pfr{^2 h_1}{\zeta^2} - \dot m_0|\nabla G|\pfr{h_1}{\zeta} = -\frac{\kappa \dot M_1^+}{\dot m_0}  + \frac{\kappa\rho_f\zeta}{\dot m_0|\nabla G|^2}\frac{\tilde D^+|\nabla G|}{\tilde Dt}+ \frac{\kappa\lambda_f}{\dot m_0}\left(\nabla\cdot\mb n -\frac{\mb n\cdot\nabla|\nabla G|}{|\nabla G|}\right).
\end{align}
The solution, which is not exponentially growing as $\zeta\to +\infty$, is given by
\begin{equation}
    h_1 = h_1(0,\mb x,t) +\frac{\kappa}{\dot m_0^2|\nabla G|}\left[\dot M_1^+ - \frac{\rho_f\lambda_f}{\dot m_0|\nabla G|}\frac{\tilde D^+|\nabla G|}{\tilde Dt} - \lambda_f\left(\nabla\cdot\mb n -\frac{\mb n\cdot\nabla|\nabla G|}{|\nabla G|}\right) \right]\zeta -\frac{\kappa\rho_f}{2\dot m_0^2|\nabla G|^3} \frac{\tilde D^+|\nabla G|}{\tilde Dt}\zeta^2, \quad \zeta>0. 
\end{equation}
From this solution, we can obtain $\pfi{h_1}{\zeta}$ at $\zeta=0^+$, 
which is to be used in the last jump condition in~\eqref{sheetjump}.

Our primary interest here is to determine the unknown function $\dot M_1^-$ in~\eqref{m1 equation}. To this end, it is sufficient to integrate~\eqref{temp1}-\eqref{enthal1} over the entire preheat zone, i.e., from $\zeta=-\infty$ to $\zeta=0^-$.  The integration is subject to the  requirement that $\theta_1$ and $h_1$ and their gradients vanish as $\zeta\to-\infty$ and the jump conditions~\eqref{sheetjump} at the reaction sheet which are given by
\begin{align}
     \theta_1= 0 \,\, (\text{thus}\,\,\lambda_1=0), \quad & \lambda_0|\nabla G| \pfr{\theta_1}{\zeta} = \frac{\dot m_0h_1}{2} \quad \text{at} \quad \zeta=0^-,\\
     \lambda_0 |\nabla G|   \pfr{(h_1+l\theta_1)}{\zeta} &= \frac{\kappa\lambda_f\dot M_1^+}{\dot m_0^2} - \frac{\kappa\rho_f\lambda_f^2}{\dot m_0^3|\nabla G|}\frac{\tilde D^+|\nabla G|}{\tilde Dt} - \frac{\kappa\lambda_f^2}{\dot m_0^2}\left(\nabla\cdot\mb n -\frac{\mb n\cdot\nabla|\nabla G|}{|\nabla G|}\right) \quad \text{at} \quad \zeta=0^-.
\end{align} 
Performing the required integration in the preheat zone then yields
\begin{align}
      \frac{1}{|\nabla G|}\int_{-\infty}^{0^-}RHS_{\theta_1}^-d\zeta &= \frac{\dot m_0 h_1(0,\mb x,t)}{2}, \\  \frac{1}{|\nabla G|}\int_{-\infty}^{0^-}[RHS_{h_1}^-+\kappa\theta_1]d\zeta &=-\dot m_0 h_1(0,\mb x,t) + \frac{\kappa\lambda_f\dot M_1^+}{\dot m_0^2} - \frac{\kappa\rho_f\lambda_f^2}{\dot m_0^3|\nabla G|}\frac{\tilde D^+|\nabla G|}{\tilde Dt} - \frac{\kappa\lambda_f^2}{\dot m_0^2}\left(\nabla\cdot\mb n -\frac{\mb n\cdot\nabla|\nabla G|}{|\nabla G|}\right),
\end{align}
which upon eliminating $h_1(0,\mb x,t)$ provides the desired relation which determines $\dot M_1^-$,
\begin{align}
     \frac{1}{|\nabla G|}\int_{-\infty}^{0^-} \left(RHS_{h_1}^-+2 RHS_{\theta_1}^-  +\kappa\theta_1\right)d\zeta &=\frac{\kappa\lambda_f\dot M_1^+}{\dot m_0^2} - \frac{\kappa\rho_f\lambda_f^2}{\dot m_0^3|\nabla G|}\frac{\tilde D^+|\nabla G|}{\tilde Dt} - \frac{\kappa\lambda_f^2}{\dot m_0^2}\left(\nabla\cdot\mb n -\frac{\mb n\cdot\nabla|\nabla G|}{|\nabla G|}\right). \label{M1 equation}
\end{align}
The first two terms on the left-hand side can be evaluated by integrating~\eqref{RHStheta1} and~\eqref{RHSh1} across the preheat zone, which yields
\begin{align}
    \frac{1}{|\nabla G|}\int_{-\infty}^{0^-}RHS_{\theta_1}^-d\zeta&=\dot M_1^- + \frac{1}{\dot m_0|\nabla G|}\left\{\mc I_1'\frac{\tilde D^-|\nabla G|}{\tilde Dt} - \dot m_0\mc I_1'\nabla^2 G  + \mc I_2'\nabla\cdot(|\nabla G|\mathbfcal P\mb V_0^-) +  \mc I_3'\nabla\cdot(|\nabla G|\mathbfcal P\mb g)\right\}, \label{tempflux} \\
    \frac{1}{|\nabla G|}\int_{-\infty}^{0^-}RHS_{h_1}^-d\zeta &= \dot M_1^- h_f + \frac{1}{\dot m_0|\nabla G|}\left\{\mc I_1''\frac{\tilde D^-|\nabla G|}{\tilde Dt} - \dot m_0\mc I_1''\nabla^2G  + \mc I_2''\nabla\cdot(|\nabla G|\mathbfcal P\mb V_0^-) +  \mc I_3''\nabla\cdot(|\nabla G|\mathbfcal P\mb g)\right\} \nonumber \\
    & \quad -\frac{\kappa \mc I_4''}{\dot m_0^2} \frac{\nabla^2G}{|\nabla G|},\label{enthalpyflux}
\end{align}
where 
\begin{align}
    \mc I_1' &= \int_0^1 \frac{\lambda_0}{\theta_0}[1-\rho_0(1-\theta_0)] d\theta_0, \quad
    \mc I_2' = \int_0^1 \frac{\rho_0\lambda_0}{\mu_0\theta_0}(\mu_0-1)(1-\theta_0)  d\theta_0, \quad
  \mc I_3' = \int_0^{1} \frac{\rho_0\lambda_0}{\mu_0\theta_0}(1-\rho_0)(1-\theta_0) d\theta_0 \,,\\
  \mc I_1'' &= \int_0^1 \frac{\lambda_0}{\theta_0}[ h_f-\rho_0(h_f-h_0)] d\theta_0, \quad
    \mc I_2'' = \int_0^1 \frac{\rho_0\lambda_0}{\mu_0\theta_0}(\mu_0-1)(h_f-h_0)  d\theta_0, \\
  \mc I_3'' &= \int_0^{1} \frac{\rho_0\lambda_0}{\mu_0\theta_0}(1-\rho_0)(h_f-h_0) d\theta_0 \,, \quad \mc I_4'' = \int_0^1 \int_0^{\theta_0}\lambda_0'd\theta_0'\,\frac{\lambda_0}{\theta_0}d\theta_0. 
\end{align}
To determine the third term on the left side of~\eqref{M1 equation}, we can use the formal solution for $\theta_1$ provided in~\eqref{theta1sol} and do an integration by parts.  Alternatively, by rewriting equation~\eqref{temp1} as $|\nabla G|^2\pfi{^2(\lambda_0\theta_1)}{\zeta^2}-\dot m_0|\nabla G|\pfi{\theta_1}{\zeta}=RHS_{\theta_1}$  using the relation $\lambda_1=\theta_1d\lambda_0/d\theta_0$ and integrating twice across the preheat zone, we find
\begin{align}
    \int_{-\infty}^{0}\theta_1 d\zeta =-\frac{1}{\dot m_0|\nabla G|}\int_{-\infty}^0 \int_{-\infty}^{\zeta}\left[RHS_{\theta_1} \right]_{\zeta'} d\zeta'd\zeta = \frac{1}{\dot m_0|\nabla G|}\int_{-\infty}^0 \zeta RHS_{\theta_1}  d\zeta  \label{theta_1integral}    
\end{align}
where the second equality comes about by swapping the integration order and integrating once. To explicitly evaluate this integral, it is convenient to use the middle expression while integrating the first and last terms in~\eqref{RHStheta1} and the last expression while integrating the remaining three terms. After some lengthy calculations involving integration by parts and the usage of the relation $\zeta=-L|\nabla G|/\dot m_0$ with $L(\theta_0)=\int_{\theta_0}^1d\theta_0'\lambda_0'/\theta_0'$, we obtain
\begin{align}
    \frac{\kappa}{|\nabla G|}\int_{-\infty}^{0}\theta_1d\zeta &=  - \frac{\kappa \dot M_1^-}{\dot m_0^2}\int_0^1\lambda_0d\theta_0 - \frac{\kappa}{\dot m_0^3|\nabla G|}\left\{\mc I_1'''\frac{\tilde D^-|\nabla G|}{\tilde Dt} - \dot m_0\mc I_0'''\nabla^2G  + \mc I_2'''\nabla\cdot(|\nabla G|\mathbfcal P\mb V_0^-) +  \mc I_3'''\nabla\cdot(|\nabla G|\mathbfcal P\mb g)\right\} \nonumber \\
    &\quad+\frac{\kappa\mc I_4'''}{\dot m_0^2}\frac{\mb n\cdot\nabla|\nabla G|}{|\nabla G|} - \frac{\kappa\mc I_5'''}{\dot m_0^3}\frac{\mathbfcal P\mb V_0^-\cdot\nabla|\nabla G|}{|\nabla G|}  - \frac{\kappa\mc I_6'''}{\dot m_0^3}\frac{\mathbfcal P\mb g\cdot\nabla|\nabla G|}{|\nabla G|}, \label{heatflux}
\end{align}
where
\begin{align}
    &\mc I_1'''=\int_0^1[(\ln\theta_0+2)\hat{\mc I_1}+2\rho_0 L]\lambda_0 d\theta_0, \quad \mc I_0'''=\int_0^1(\hat{\mc I_1}+\rho_0 L)\lambda_0d\theta_0, \quad \mc I_2'''= \int_0^1 \left[\hat{\mc I_2}-\frac{\rho_0}{\mu_0}(\mu_0-1)L\right]\lambda_0d\theta_0,\\
    &\mc I_3'''= \int_0^1 \left[\hat{\mc I_3}-\frac{\rho_0}{\mu_0}(1-\rho_0)L\right]\lambda_0d\theta_0, \quad \mc I_4''' = \int_0^1[(\ln\theta_0+1)\hat{\mc I_1}+\rho_0 L - \lambda_0\ln\theta_0]\lambda_0d\theta_0, \\
    &\mc I_5''' = \int_0^1 \left[\hat{\mc I_2}(\ln\theta_0+1)-\frac{\rho_0}{\mu_0}(\mu_0-1)L\right]\lambda_0d\theta_0, \quad \mc I_6'''= \int_0^1 \left[\hat{\mc I_3}(\ln\theta_0+1)-\frac{\rho_0}{\mu_0}(1-\rho_0)L\right]\lambda_0d\theta_0.
\end{align}
Finally, we can express the right-hand side of equation~\eqref{M1 equation} in terms of minus variables using~\eqref{Dplusminus} and~\eqref{M1plusminus}, so that it becomes
\begin{align}
   &\frac{\kappa\lambda_f}{\dot m_0^2}\left\{\dot M_1^- + \frac{(\mc I_1-\rho_f\lambda_f)}{\dot m_0|\nabla G|}\frac{\tilde D^-|\nabla G|}{\tilde Dt} -(\mc I_1+\lambda_f)\frac{\nabla^2 G}{|\nabla G|} + \lambda_f(1+\rho_f) \frac{\mb n\cdot\nabla|\nabla G|}{|\nabla G|} + \frac{\mc I_2\nabla\cdot [|\nabla G|\mathbfcal P \mb V_0^-]}{\dot m_0|\nabla G|}  +\frac{\mc I_3\nabla\cdot [|\nabla G|\mathbfcal P \mb g]}{\dot m_0|\nabla G|}\right\} \nonumber \\
   & +  \frac{\kappa\rho_f\lambda_f^2}{\dot m_0^3|\nabla G|}\left( \frac{\mu_f-1}{\mu_f}\mathbfcal P\mb V_0^- + \frac{1-\rho_f}{\mu_f}\mathbfcal P\mb g \right)\cdot \nabla|\nabla G| .\label{heatfluxright}
\end{align}

Now we have all the four required expressions~\eqref{tempflux},~\eqref{enthalpyflux},~\eqref{heatflux} and~\eqref{heatfluxright} which appear in~\eqref{M1 equation}, $\dot M_1^-(\mb x,t)$ can be obtained by rearranging and combining different terms appropriately. We find
\begin{align}
    2 (1+h_f)\dot M_1^- &=  \frac{-1}{\dot m_0|\nabla G|}\frac{\tilde D^-|\nabla G|}{Dt}\left[\mc I_1'' + 2 \mc I_1' - \frac{\kappa\lambda_f}{\dot m_0^2}(\mc I_1''' + \mc I_1 -\rho_f \lambda_f)  \right] + \frac{\nabla^2G}{|\nabla G|}\left[\mc I_1'' + 2\mc I_1'- \frac{\kappa}{\dot m_0^2}\left(\mc I_0'''-\mc I_4''+\lambda_f(\mc I_1+\lambda_f)\right)\right] \nonumber \\
    &\quad  - \frac{\nabla\cdot [|\nabla G|\mathbfcal P \mb V_0^-]}{\dot m_0|\nabla G|} \left[\mc I_2' + \mc I_2'' - \frac{\kappa}{\dot m_0^2}(\mc I_2'''+\lambda_f \mc I_2)\right] - \frac{\nabla\cdot [|\nabla G|\mathbfcal P \mb g^-]}{\dot m_0|\nabla G|} \left[\mc I_3' + \mc I_3''- \frac{\kappa}{\dot m_0^2}(\mc I_3'''+\lambda_f \mc I_3)\right]\nonumber \\
    &\quad  -\frac{\mb n\cdot \nabla|\nabla G|}{|\nabla G|}\frac{\kappa}{\dot m_0^2}[\mc I_4'''-\lambda_f^2(1+\rho_f)]  + \frac{\kappa}{\dot m_0^3}\frac{\mathbfcal P\mb V_0^-\cdot\nabla|\nabla G|}{|\nabla G|}\left[\mc I_5''' + \frac{\rho_f\lambda_f^2}{\mu_f}(\mu_f-1)\right] \nonumber \\
    &\quad  + \frac{\kappa}{\dot m_0^3}\frac{\mathbfcal P\mb g\cdot\nabla|\nabla G|}{|\nabla G|}\left[\mc I_6''' + \frac{\rho_f\lambda_f^2}{\mu_f}(\mu_f-1)\right].
\end{align}
This expression can be further simplified using the following vector identities
\begin{align}
    &\frac{1}{|\nabla G|}\frac{\tilde D^-|\nabla G|}{\tilde Dt} = \dot m_0 \frac{\mb n \cdot \nabla |\nabla G|}{|\nabla G|} -\mb n\mb n : \nabla \mb V_0^- , \qquad \frac{\nabla^2 G}{|\nabla G|}=  \frac{\mb n \cdot \nabla |\nabla G|}{|\nabla G|} + \nabla\cdot \mb n, \nonumber\\
    &\frac{\nabla\cdot (|\nabla G| \mathbfcal P \mb A)}{|\nabla G|}=   \nabla_t\cdot(\mathbfcal P \mb A) \quad \text{for an arbitrary solenoidal vector } \mb A,  \nonumber
\end{align}
where $\nabla_t\cdot(\mathbfcal P \mb A)=-(\mb A\cdot \mb n)\nabla\cdot\mb n-\mb n\mb n:\nabla \mb A$ is the surface divergence of the tangential component $\mb A_t = \mathbfcal P \mb A$ of the solenoidal vector $\mb A$.

\subsection{First correction to the flow field}

With the leading-order velocity $\mb v_0$ determined in~\eqref{innerflow}, the first correction  to the pressure field, $p_1$, is determined by integrating~\eqref{leadingmomentum} with respect to $\zeta$. We find
\begin{align} \label{p1sol}
    p_1 (\zeta,\mb x,t) =  \begin{cases}
        \displaystyle P_1^-(\mb x,t) - \hat\Gamma_1   - \frac{\mb V_0^-\cdot \mb n}{\dot m_0}\hat\Gamma_2  - \frac{\mb g\cdot \mb n}{\dot m_0}\hat\Gamma_3, \, 
        & \zeta < 0\\
        \displaystyle P_1^+(\mb x,t) \,, 
        & \zeta > 0,     
    \end{cases}
\end{align}
where
\begin{align}
    &\hat \Gamma_1(\theta_0) = \int_0^{\theta_0} \frac{\mu_0'\lambda_0'}{\rho_0'\theta_0'}(1-\rho_0')d\theta_0', \qquad \hat\Gamma_2(\theta_0) = \int_0^{\theta_0}\frac{\lambda_0'}{\theta_0'}(\mu_0'-1)d\theta_0', \qquad \hat\Gamma_3(\theta_0) = \int_0^{\theta_0}\frac{\lambda_0'}{\theta_0'}(1-\rho_0')d\theta_0'. \\
    &P_1^+(\mb x,t) = P_1^-(\mb x,t) - \Gamma_1   - \frac{\mb V_0^-\cdot \mb n}{\dot m_0}\Gamma_2  - \frac{\mb g\cdot \mb n}{\dot m_0}\Gamma_3, \qquad \text{with} \qquad \Gamma_i\equiv\hat\Gamma_i(1). 
\end{align}

Next, we can introduce the auxiliary first-order outer problems 
\begin{align}
    \nabla \cdot \mb V_1^- =0, \qquad -\mb V_1^- = \nabla P_1^- \quad \text{as} \quad \zeta \to -\infty,\\
    \nabla \cdot \mb V_1^+ =0, \qquad -\mu_f\mb V_1^+ = \nabla P_1^+ \quad \text{as} \quad \zeta \to +\infty.
\end{align}
along with the kinematic constraints
\begin{equation}
    \mb V_1^-\cdot \mb n = \dot M_1^- \qquad \text{and} \qquad \rho_f\mb V_1^+\cdot \mb n = \dot M_1^+, \quad \text{on either side of the flame, respectively.}
\end{equation}
The full composite velocity field $\mb v_1$ is determined in a manner similar  to that of $\mb v_0$ discussed in Section~\ref{sec:leadingorderflow}. The normal component of $\mb v_1$ is determined through the first-order G-equation,
\begin{equation}
    \rho_0 \mb v_1 \cdot \nabla G + \frac{\rho_1}{\rho_0}\dot m_0 |\nabla G| = \dot m_1 |\nabla G| \quad \Rightarrow \quad \rho_0 \mb v_1\cdot \mb n = \dot m_1- \frac{\rho_1}{\rho_0}\dot m_0.
\end{equation}
On the other hand, the tangential component of $\mb v_1$ is obtained from the second-order momentum equation,
\begin{equation}
    -\mu_0 \mb v_1 - \mu_1\mb v_0 = \nabla p_1 - \rho_1 \mb g + \nabla G \pfr{p_2}{\zeta} , \qquad \Rightarrow \qquad -\mu_0 \mathbfcal P \mb v_1 - \mu_1\mathbfcal P\mb v_0 = \mathbfcal P\nabla p_1 - \rho_1 \mathbfcal P\mb g. \label{secondmomentum}
\end{equation}
Combining the two components of $\mb v_1$, we obtain
\begin{align}
    \mb v_1 &= \frac{1}{\rho_0}\left(\dot m_1-\frac{\rho_1}{\rho_0}\dot m_0\right)\mb n + \frac{1}{\mu_0}\mathbfcal P \mb V_1^- - \frac{\mu_1}{\mu_0}\mathbfcal P\mb V_0^- + \frac{1}{\mu_0}\left[\mathbfcal P\nabla\hat\Gamma_1 + \mathbfcal P\nabla\left(\frac{\mb V_0^-\cdot \mb n}{\dot m_0}\hat\Gamma_2\right) + \mathbfcal P\nabla \left(\frac{\mb g\cdot \mb n}{\dot m_0}\hat\Gamma_3\right)\right]\nonumber \\
    &\quad + \frac{\mu_1 + \mu_0\rho_1 - \mu_1 \rho_0}{\mu_0^2}\mathbfcal P\mb g \qquad \qquad \text{for} \quad  \zeta<0  \,,\label{v1minus}\\
    \mb v_1 &= \frac{\dot M_1^+}{\rho_f}\mb n + \mathbfcal P \mb V_1^+ \qquad \qquad \text{for} \quad  \zeta>0. \label{v1plus}
\end{align}
The expressions~\eqref{p1sol}--\eqref{v1plus} provide the first-order corrections 
to the leading-order flow field. Evaluating these  across the flame 
yields 
\begin{equation}
    \llbracket P_1 \rrbracket = - \Gamma_1   - \frac{\mb V_0^-\cdot \mb n}{\dot m_0}\Gamma_2  - \frac{\mb g\cdot \mb n}{\dot m_0}\Gamma_3, \qquad \llbracket \mb V_1\cdot \mb n \rrbracket = \frac{1}{\rho_f}\dot M_1^+-\dot M_1^-.
\end{equation}
These are the first-order corrections to the continuity requirements of pressure and mass flux across the flame, arising due to the finite thicknesses of the flame.

\section{Summary of the hydrodynamic model}\label{sec:summary}

The results from the multiple-scale analysis are now used to formulate a 
closed hydrodynamic model for premixed flames under Darcy's law. In this 
formulation, the physical scales are redefined to eliminate the small parameter 
$\epsilon$ from the problem. Specifically, the flame thickness $\delta_L$ is 
adopted as the characteristic length scale, accompanied by the characteristic 
time scale $\delta_L/S_L$ and the characteristic pressure scale $\mu_u D_{T,u}/\kappa_u$. Physical properties remain non-dimensionalised by their unburned-gas values.  For clarity, the flow variables in this hydrodynamic zone are denoted using  lowercase letters, distinguishing them from the uppercase symbols used during the asymptotic derivation. These conventions are maintained throughout the remainder 
of this work.

\begin{figure}
\centering
\includegraphics[scale=0.4]{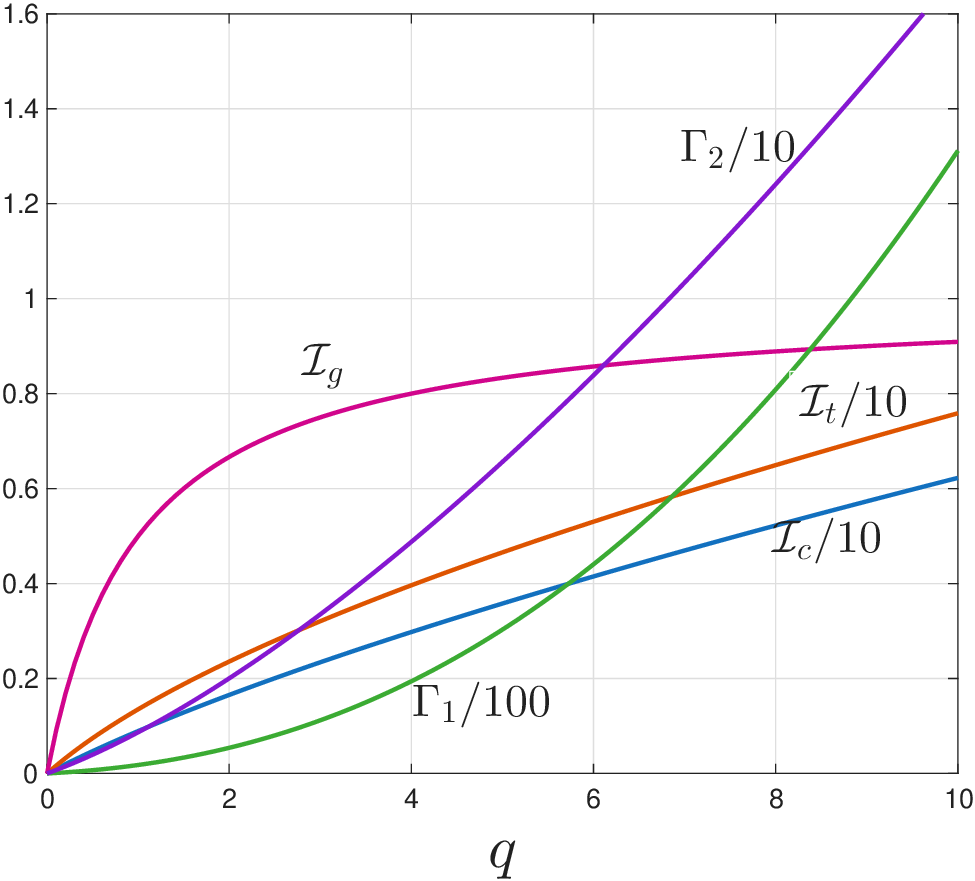} \hspace{0.5cm}
\includegraphics[scale=0.4]{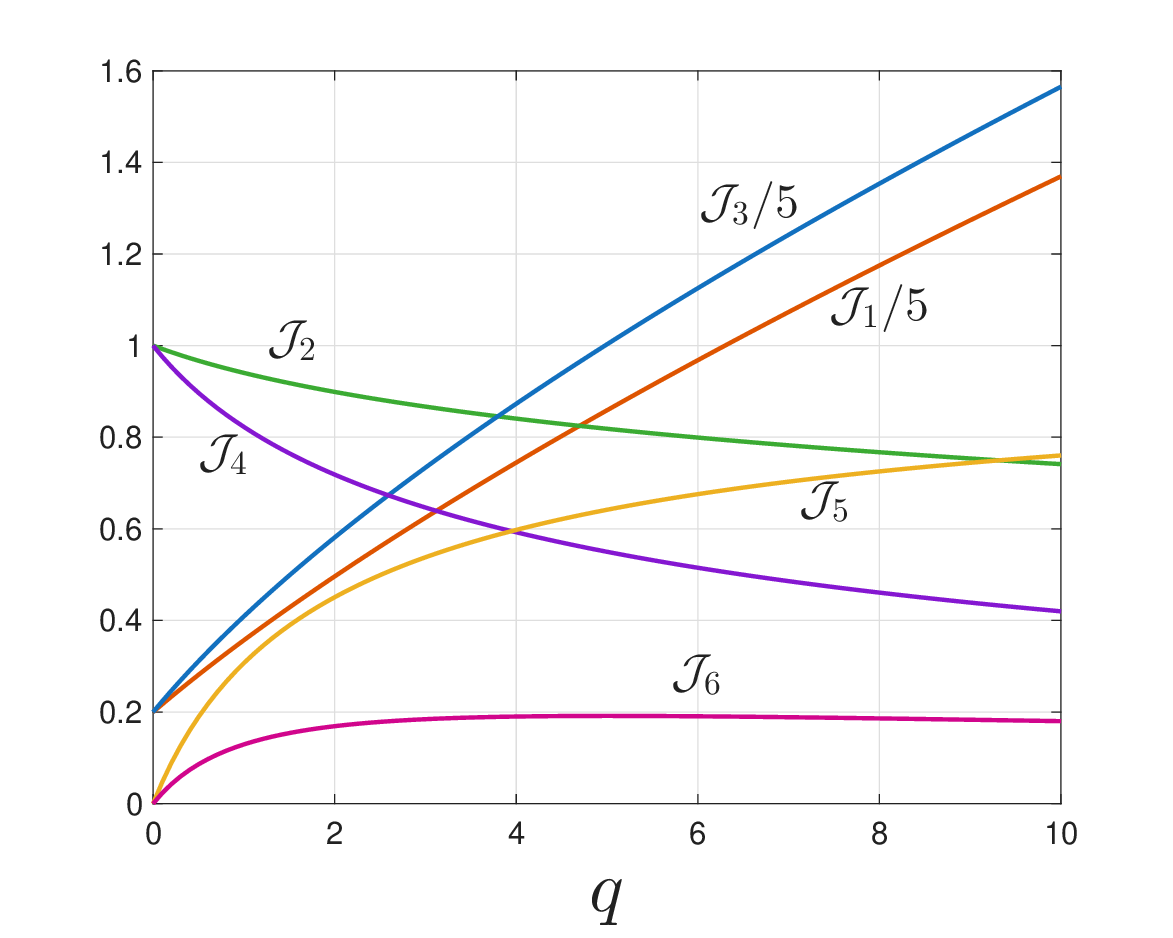}
\caption{The plot of various integrals as functions of the heat-release parameter $q$, evaluated for $\rho = 1 / (1 + q\theta)$ and $\mu = \lambda = (1 + q\theta)^{0.7}$. The left panel displays the integrals ($\Gamma_1, \Gamma_2, \Gamma_3, \mc{I}_c, \mc{I}_t, \mc{I}_g$) associated with the mass conservation and dynamic pressure jump conditions across the flame front. The right panel shows the integrals ($\mc{J}_1, \mc{J}_2, \mc{J}_3, \mc{J}_4, \mc{J}_5, \mc{J}_6$) that appear in the kinematic condition governing the local burning rate. Note that the integrals $\Gamma_3$ and $\mc{I}_c$ are mathematically identical ($\Gamma_3 = \mc{I}_c$).} 
\label{fig:integrals}
\end{figure}

In this hydrodynamic framework, the flame is treated as a discontinuity surface located at $G(\mb x,t)=0$. On either side of the flame front, $G\neq 0$, the flow is incompressible and satisfies Darcy's law,  
\begin{equation}
    \begin{cases}
        \begin{aligned}
            & \nabla \cdot \mb v^- =0, \quad \mb v^-=-\nabla p^- + \mb g,    && G < 0, \\            & \nabla \cdot \mb v^+ =0, \quad \frac{1}{\mm}\mb v^+=-\nabla p^+ + \frac{1}{\rr}\mb g , && G > 0,
        \end{aligned}
    \end{cases} \quad \Rightarrow \quad \nabla^2p^\pm=0 \quad \text{for} \quad G\neq 0,
     \label{finalproblem}
\end{equation}
where $\rr=1/\rho_f$ is the ratio of unburnt-to-burnt gas density and $\mm = 1/\mu_f$ is the ratio of unburnt-to-burnt gas viscosity (or more generally viscosity/permeability). At  the flame front, $G=0$,  the flow variables satisfy the jump conditions
 \begin{align}
    \llbracket \rho(\mb v\cdot \mb n-U_f)\rrbracket &= -\mc I_c (\dot m_0-\mb v^-\cdot \mb n)\nabla\cdot \mb n + \mc I_t \nabla_t\cdot \mb v_t^- + \mc I_g \nabla_t \cdot \mb g_t \,,  \label{interfacedarcy1}\\ \llbracket p\rrbracket &=- \Gamma_1   - \frac{\mb v^-\cdot \mb n}{\dot m_0}\Gamma_2  - \frac{\mb g\cdot \mb n}{\dot m_0}\Gamma_3  \,,  \label{interfacedarcy2}
\end{align}
where the jump operator is defined by
$\llbracket\varphi\rrbracket\equiv \varphi^+ |_{G=0^+}-\varphi^- |_{G=0^-}$.  The unit normal vector pointing toward the burned gas is $\mb n=\nabla G/|\nabla G|$, 
and $U_f = - (\pfi{G}{t})/|\nabla G|$ represents the flame-front speed along 
$\mb n$ in the laboratory frame.  Moreover, $\nabla_t\cdot \mb A_t$ denotes the surface divergence of the tangential component $\mb A_t=(\mb I-\mb n\otimes \mb n)\mb A$ of an arbitrary vector $\mb A$; when $\nabla\cdot \mb A=0$, we have $\nabla_t\cdot \mb A_t=-\mb n\mb n:\nabla \mb A-(\mb A\cdot \mb n)\nabla\cdot \mb n$. The model is further constrained by the kinematic condition
 \begin{equation} 
\pfr{G}{t}+\mb v^-\cdot \nabla G=\dot m^- |\nabla G|,   \label{kinematic}
\end{equation}
applied on the unburnt gas side $G=0^-$. The burning rate  $\dot m^-$ at the unburnt gas side, is found to be  
\begin{align}
    \dot m^- &= \dot m_0 +\mc M_c^{(1)} \dot m_0 \nabla \cdot \mb n -\mc M_c^{(2)} (\mb v^-\cdot \mb n)\nabla\cdot \mb n - \mc M_t \nabla_t\cdot \mb v_t^- - \mc M_g \nabla_t \cdot \mb g_t \nonumber \\ & \quad - \mc N_n \dot m_0 \frac{\mb n\cdot \nabla|\nabla G|}{|\nabla G|} + \mc N_{v} \frac{\mb v_t^-\cdot \nabla|\nabla G|}{|\nabla G|} + \mc N_{g} \frac{\mb g_t\cdot \nabla|\nabla G|}{|\nabla G|},\label{m1final}
\end{align}
where $\dot m_0$ is the burning-rate of non-adiabtic planar flame. Here, $\{\mc M_c^{(1)},\mc M_c^{(2)}\}$ 
denote the \textit{curvature Markstein numbers}, $\mc M_t$ is the \textit{tangential flow-strain Markstein number}, 
and $\mc M_g$ is a novel parameter specific to Darcy flows that may be termed the \textit{gravity-strain Markstein number}. The remaining parameters $\mc N_n$, $\mc N_v$, 
and $\mc N_g$ are coefficients which capture the effects of spatial variations in the level-set gradient induced by  heat loss.  These parameters are given by
\begin{align}
    \mc M_c^{(1)} &= \frac{1}{\dot m_0(1+h_f)}\left\{\mathcal{J}_1 + \frac{l}{2}\mathcal{J}_2 - \frac{\kappa}{2\dot m_0^2} \mc J_7\right\}, \qquad \mc M_c^{(2)} =  \frac{1}{\dot m_0(1+h_f)}\left\{\mathcal{J}_1 + \frac{l}{2}\mathcal{J}_2 - \frac{\kappa}{2\dot m_0^2} \mc J_8\right\},\label{Mc} \\   
    \mc M_t &=  \frac{1}{\dot m_0(1+h_f)}\left\{\mathcal{J}_3 + \frac{l}{2}\mathcal{J}_4 - \frac{\kappa}{2\dot m_0^2} \mc J_9\right\}, \qquad
   \mathcal{M}_g =  \frac{1}{\dot m_0(1+h_f)}\left\{\mathcal{J}_5 - \frac{l}{2}\mathcal{J}_6 - \frac{\kappa}{2\dot m_0^2} \mc J_{10}\right\},\label{MtMg}\\
   \mc N_n & =  \frac{1}{\dot m_0(1+h_f)}\frac{\kappa}{2\dot m_0^2}\mc J_{11}, \qquad \mc N_v =  \frac{1}{\dot m_0(1+h_f)}\frac{\kappa}{2\dot m_0^2}\mc J_{12}, \qquad \mc N_g =   \frac{1}{\dot m_0(1+h_f)}\frac{\kappa}{2\dot m_0^2}\mc J_{13}. \label{N}
\end{align}
The various integrals $\mc I_i$, $\Gamma_i$ and $\mc J_i$, appearing in the above hydrodynamic model are defined by the equation of state $\rho=\rho(\theta)$ and the temperature dependences of the transport coefficients $\mu=\mu(\theta)$ and $\lambda=\lambda(\theta)$, with $\theta\in[0,1]$ being the non-dimensional temperature. Specifically, we have
\begin{align}
&\mc I_c = \int_0^1 \frac{\lambda}{\theta}(1-\rho)d\theta, \qquad \mc I_t = \int_0^1 \frac{\lambda}{\theta}\left(1-\frac{\rho}{\mu}\right)d\theta, \qquad \mc I_g = \int_0^1 \frac{\rho\lambda}{\mu\theta}(1-\rho)d\theta, \\
&\Gamma_1 = \int_0^1 \frac{\mu\lambda}{\rho\theta}(1-\rho)d\theta, \qquad \Gamma_2 = \int_0^1\frac{\lambda}{\theta}(\mu-1)d\theta, \qquad \Gamma_3 = \int_0^1\frac{\lambda}{\theta}(1-\rho)d\theta,\\ 
    &\mc J_1 = \int_0^1 \frac{\lambda}{\theta}[1-\rho(1-\theta)] d\theta, \qquad
    \mc J_2 =  -\int_0^1\rho\lambda \ln\theta d\theta, \qquad
  \mc J_3 = \int_0^1 \frac{\lambda}{\theta}\left[1-\frac{\rho}{\mu}(1-\theta)\right] d\theta,\\
  &\mc J_4 =  - \int_0^1\frac{\rho}{\mu}\lambda \ln\theta d\theta,\qquad
  \mc J_5 = \int_0^{1} \frac{\rho\lambda}{\mu\theta}(1-\rho)(1-\theta) d\theta,\qquad
  \mc J_6 =  -\int_0^1\frac{\rho\lambda}{\mu}(1-\rho)\ln\theta d\theta,
\end{align}
which are plotted in Fig.~\ref{fig:integrals} as functions of $q$ for $\rho=1/(1+q\theta)$ and $\mu=\lambda=(1+q\theta)^{0.7}$. For brevity, the expressions for $\mc J_7$ to $\mc J_{13}$, which are not needed in the absence of heat loss, are relegated to Appendix A. 
Finally, the planar burning 
rate parameters are governed by
\begin{equation}
    h_f = \ln \dot m_0^2, \qquad \dot m_0^2\ln\dot m_0^2 = - \frac{\kappa}{e\kappa_\ext}, \quad \kappa_\ext= \frac{1}{e\left(\lambda_f+\int_0^1\lambda_0 d\theta_0 \right)}.
\end{equation}

\textbf{Remark on the burnt-gas heat-loss zone:} On the burnt gas side of the flame, the gas temperature will eventually cool down to cold temperature due to heat loss. However, in the large activation energy analysis, this zone, referred as the \textit{convection-heat loss zone}~\cite{williams2018combustion}, occurs on an extremely large length scale, $x\sim \beta$ (dimensionally of order $\beta\delta_L$). The multiple scale analysis of the previous section assumes a hydrodynamic scale of order $x\sim 1/\ep$ (dimensionally of order $L$) with the assumption that $\beta^{-1}\ll \ep$. Thus, our model is valid when $\beta \delta_L\gg L$, i.e., the convection-heat loss zone is much larger than the hydrodynamic zone. As a result, the burnt-gas temperature in the hydrodynamic zone is still the adiabatic flame temperature, in the first approximation. This is similar to the approach adopted in~\cite{clavin1985effect,keller1994transient,matalon2009multi}.

\subsection{The adiabatic limit ($\kappa=0$)}

Admittedly, the results summarised above are quite complex. However, this level 
of complexity is to be expected from similar asymptotic studies in the 
literature based on the Navier--Stokes equations, 
especially when heat losses are included~\cite{keller1994transient,matalon2009multi}. It is instructive to point out 
that the complexity lies primarily in the formula for the burning rate $\dot m^-$, given in~\eqref{m1final}, through which both heat-loss and 
non-unity Lewis-number effects enter the hydrodynamic model. Specifically, the difficulty stems from the combustion equations governing 
temperature and fuel mass fraction (or equivalently enthalpy), rather than 
the hydrodynamic equations of continuity and momentum.
 Fortunately, significant simplifications 
occur if the adiabatic assumption is adopted ($\kappa=0$).

Indeed, when heat losses are neglected by setting $\kappa=0$ so that $\dot m_0=1$, 
the burning-rate formula simplifies to 
\begin{equation}
   \dot m^- = 1 +\mc M_c (1-\mb v^-\cdot \mb n)\nabla\cdot \mb n - \mc M_t \nabla_t\cdot \mb v_t^- - \mc M_g \nabla_t \cdot \mb g_t, \label{adiabaticdotformula}
\end{equation}
with the  Markstein numbers given by
\begin{equation}
    \mc M_c =  \mathcal{J}_1 + \frac{l}{2}\mathcal{J}_2, \qquad \mc M_t = \mathcal{J}_3 + \frac{l}{2}\mathcal{J}_4, \qquad \mc M_g = \mathcal{J}_5 - \frac{l}{2}\mathcal{J}_6 \,. \label{Mckzero}
\end{equation}
The implications of this simplified version, applicable under the adiabatic 
approximation, were used within simple canonical configurations in our recent 
work~\cite{rajamanickam2026flame}. However, that study did not provide a 
derivation for these expressions, which is now detailed herein. Crucially, we note that 
\begin{equation}
    \mc M_t \gtrless \mc M_c \qquad \text{if} \qquad l \lessgtr l_*  \equiv 2\frac{\mc J_3-\mc J_1}{\mc J_2-\mc J_4} \,.\label{lstar}
\end{equation}
The critical value $l_*$ of the reduced Lewis number obeys the inequality $l_*\geq 4$  or equivalently $Le_*\geq 1+4/\beta$. This suggests that the tangential strain contribution is typically larger than the curvature contribution ($\mc M_t\geq \mc M_c$), except for large Lewis number mixtures. 

The expressions~\eqref{adiabaticdotformula}-\eqref{Mckzero} may be compared with the classical Clavin--Williams formula derived for a one-step chemistry model under Navier--Stokes hydrodynamics~\cite{clavin1982effects}, which in the present notation reads
\begin{equation}
    \dot m^-=1+ \mc M\left\{(1-\mb v^-\cdot \mb n)\nabla\cdot\mb n-\nabla_t\cdot\mb v_t^-\right\} = 1-\mc M\mathbb K, \qquad \text{where}\qquad \mc M=\mc J_1 + \frac{l}{2}\mc J_2,
\end{equation}
and $\mathbb K=-\nabla\cdot\mb n-\mb n\otimes\mb n :\nabla \mb v^-$ is the kinematic flame stretch. Notably, if we assume a constant viscous resistance by setting $\mu=1$, our curvature and tangential strain integrals collapse such that $\mathcal{J}_3 = \mathcal{J}_1$ and $\mathcal{J}_4 = \mathcal{J}_2$. Under this condition, the distinct Markstein numbers coincide ($\mc M_c = \mc M_t = \mc M$), and the curvature and flow-strain terms in our model combine to exactly reproduce the classical stretch term $-\mc M \mathbb K$. However, the gravity-induced strain term $\mc M_g \nabla_t \cdot \mb g_t$ persists even when $\mu=1$, highlighting a fundamental distinction introduced by the Darcy-law momentum balance formulation.

\subsection{Discussions of the interfacial conditions}\label{sec:discussions}

The three interfacial conditions~\eqref{interfacedarcy1}--\eqref{kinematic} 
correspond to mass conservation, the dynamic condition, and the kinematic condition. 
All three relations now account for the finite-thickness effects of the flame. 

The mass conservation jump~\eqref{interfacedarcy1} can be compared with its Navier--Stokes 
counterpart~\cite{matalon2003hydrodynamic}, namely
\begin{equation}
     \llbracket \rho(\mb v\cdot \mb n-U_f)\rrbracket = -\mc I_c \left[(\dot m_0-\mb v^-\cdot \mb n)\nabla\cdot \mb n + \nabla_t\cdot \mb v_t^-\right] = \mc I_c \mathbb K \,,
\end{equation}
where $\mathbb K = -\dot m_0 \nabla \cdot \mb n - \mb n \mb n : \nabla \mb v^-$ 
is the kinematic flame stretch. Thus, under Darcy's law, flame responds differently to curvature and tangential straining, resulting in two unequal coefficients $\mc I_c \neq \mc I_t$. This inequality stems from viscosity variations coupled with a jump in the 
tangential velocity; when $\mu=1$, we also have $\mc I_c=\mc I_t$. Moreover, the tangential straining by the gravity introduces additional contributions to the jump condition, which has no analogue in the Navier--Stokes formulation. 

The dynamic condition~\eqref{interfacedarcy2} for the pressure jump cannot be 
directly compared with Navier--Stokes formulations, as the two frameworks 
originate from completely different momentum balance laws. Under Darcy's law, 
the pressure jump $\llbracket p \rrbracket$ across the flame thickness is 
driven by three distinct physical contributions. The first term $-\Gamma_1<0$  in~\eqref{interfacedarcy2} corresponds to a pressure drop as we go from unburnt gas side to burnt gas side. This pressure drop is simply associated with the cost of flow accelerating along $\mb n$ through the flame due to thermal expansion and exists even when $\mb v^-=0$ (stationary unburnt gas) or $\mu=1$ (constant viscosity). The second term $-\mb v^-\cdot \mb n \Gamma_2/\dot m_0$ is associated with viscosity variations ($\mu \neq 1$) and the unburnt gas flow. When $\mb v^-\cdot \mb n>0$ (unburnt gas flows into the flame), flow moves from low to high viscosity, increasing the pressure drop across the flame. When $\mb v^-\cdot \mb n<0$ (unburnt gas flows away, so burnt gas flows into the flame), flow moves from high to low viscosity, reducing the pressure drop. The third term $-\mb g\cdot n \Gamma_2/\dot m_0$ is a gravitational correction due to density variation across the flame. When $\mb g\cdot \mb n>0$ (gravity points towards burnt gas), the density drop reduces the gravitational body force in the burnt gas, adding to the pressure drop. When $\mb g\cdot \mb n<0$ (gravity points towards unburnt gas), the opposite occurs, reducing the pressure drop. For illustration, if we consider a planar flame propagating in the presence of a uniform flow $\mc V$ with $\mc V>0$ pertaining to  flow opposing the flame propagation and $\mc V<0$ to flow aiding the flame propagation, we have
\begin{equation} \label{pjump}
    \llbracket p \rrbracket = -\Gamma_1 - \frac{\mc V}{\dot m_0} \Gamma_2 - \frac{\mb g\cdot \mb n}{\dot m_0}\Gamma_3 \,.
\end{equation}

The kinematic condition~\eqref{kinematic} involves the local burning rate 
$\dot m^-$, which is substantially affected by heat loss, non-unity Lewis numbers, 
and novel geometric terms specific to Darcy's law. In the presence of heat 
loss, it becomes necessary to introduce two distinct curvature Markstein 
numbers, $\mc M_c^{(1)}$ and $\mc M_c^{(2)}$; the first multiplies the curvature 
term $\nabla \cdot \mb n$, while the second multiplies the coupled term 
$(\mb v^- \cdot \mb n)\nabla\cdot \mb n$. These two parameters coincide in the 
adiabatic limit ($\kappa = 0$), as shown in~\eqref{Mckzero}.  The model also yields a tangential flow-strain Markstein number $\mc M_t$ and a 
gravity-strain Markstein number $\mc M_g$. Crucially, under Darcy's law, $\mc M_t$ 
remains distinct from $\mc M_c$, and the gravity term $\mc M_g$ emerges as a 
completely unique feature of the porous medium framework. These fundamental 
distinctions hold true regardless of the presence or absence of heat losses.

\section{Stability of planar flames in the presence of an imposed flow and gravity} \label{sec:stability}

The stability analysis of planar premixed flames under Darcy's law in the presence of an imposed flow and gravity has been investigated in~\cite{daou2025hydrodynamic}. However, this analysis was based on the Markstein model within which the finite-thickness corrections to the interfacial conditions are ignored; see e.g.~\cite{creta2011strain,matalon2018darrieus}. We shall revisit this problem here accounting now for the aforementioned corrections.

Consider a two-dimensional planar flame propagating in the negative $y$-direction in the presence of a uniform oncoming flow of non-dimensional magnitude $\mc V$ (measured with $S_L$); the flow opposes the flame when $\mc V>0$ and aids it when $\mc V<0$. The non-dimensional gravity vector is taken as $\mb g = \mathrm{g} \mb e_y$, where $\mathrm{g} < 0$ when $\mb g$ points toward the unburnt gas and $\mathrm{g} > 0$ otherwise. The undisturbed flame front is planar and is located at $G = y+(\dot m_0-\mc V)t=0$; the flame is thus stationary when $\mc V=\dot m_0$, following the configuration of Joulin \& Sivashinsky~\cite{joulin1994influence}. In the frame moving with planar flame, which is now located at $y=0$ in the new coordinate, the base state flow field is given by
\begin{align}
    y<0: &\,\, \mb v=\dot m_0\mb e_y, \quad p = (\mathrm{g} - \mc V) y, \label{base3} \\
    y>0: &\,\, \mb v = \rr \dot m_0 \mb e_y, \quad p = \left[ \frac{\gr}{\rr} - \frac{\mc V+(\rr-1)\dot m_0}{\mm} \right] y - \Gamma_1 -\Gamma_2 - \frac{\gr}{\dot m_0}\Gamma_3 \,, \label{base4}
\end{align}
where use has been made of the jump conditions, in particular of~\eqref{pjump}.

We have carried out a linear stability analysis of this planar flame solution 
following the methodology we used in~\cite{daou2025hydrodynamic}. The primary 
outcome of our analysis is the dispersion relation $s = s(k)$, which links 
the perturbation growth rate $s$ to the wavenumber $k$: 
\begin{equation}
    s = \frac{ak - bk^2 - dk^3}{1 + ck} \,, \qquad k > 0 \,, \label{disp}
\end{equation}
where the coefficients are given by
\begin{align}
a &=  \frac{\rr-1}{1+\mm} + \frac{1-\mm}{1+\mm}\mc V + \frac{\rr-1}{1+\mm}\frac{\mm}{\rr}\mathrm{g},\qquad c = \frac{\rr-1}{1+\mm}\mc M_t + \frac{\mm\Gamma_2/\dot m_0 - \rr\mc I_t}{1+\mm}, \\
b&= \frac{\rr+\mm}{1+\mm} \left[\dot m_0(\mc M_t + \mc M_c^{(1)}-\mc M_c^{(2)})+\gr \mc M_g \right] + a \mc M_t  - \frac{\rr}{1+\mm} \left( \dot m_0\mc I_t + \mathrm{g} \mc I_g \right),\\    
    d&= \frac{\mm}{1+\mm}\Gamma_2\left(\mc M_t + \frac{\mathrm{g}}{\dot m_0} \mc M_g\right) + \frac{\rr}{1+\mm}\mathrm{g} \left(\mc I_g\mc M_t-\mc I_t \mc M_g \right) + \frac{\mm \Gamma_2 - \dot m_0 \rr \mc I_t}{1+\mm}\left( \mc M_c^{(1)}-\mc M_c^{(2)}\right).
\end{align}
The new dispersion relation~\eqref{disp}, derived under Darcy’s law, can be compared directly to the classical Clavin--Garcia dispersion relation derived from the Navier--Stokes equations. A notable feature of our current dispersion relation is the presence of the cubic term $-dk^3$, which arises 
as a direct consequence of implementing higher-order corrections in the mass 
conservation and dynamic jump conditions. The significance of such cubic 
terms was recently highlighted by Bechtold and Matalon~\cite{bechtold2026long}, 
who worked out even higher-order corrections than those found in the classical 
Clavin--Garcia model or our present formulation. However, the asymptotic consistency 
of models incorporating cubic terms remains a largely unexplored open question in flame theory 
that deserves further attention.
 
Near the instability threshold $a \to 0^+$ (where $k \sim a$ and $s \sim a^2$),  the dispersion relation simplifies to 
\begin{equation}
    s = a|k| - b k^2  \quad \text{with} \quad b = \frac{\rr+\mm}{1+\mm}\mc M_t + \frac{\rr+\mm}{1+\mm}\mathrm{g} \mc M_g - \frac{\rr}{1+\mm}(\mc I_t + \mathrm{g} \mc I_g) \sim O(1).
\end{equation}
As a simple implication of this result, and following the approach of Sivashinsky~\cite{ashinsky1988nonlinear,michelson1977nonlinear}, one obtains a Michelson--Sivashinsky equation for the flame shape deviation $f(x,t)$ from its unperturbed planar value $y=(\mc V-1)t$, namely
\begin{equation}
f_t + \tfrac{1}{2}f_x^2 = a \mc H(f_x) + b f_{xx}, \label{MS}
\end{equation}
where $\mathcal{H}$ denotes the Hilbert transform, defined in Fourier space as $\mathcal{F}\{\mathcal{H}(f)\}(k) = -i\operatorname{sgn}(k)\hat{f}(k)$, so that $\mathcal{F}\{\mathcal{H}(f_x)\}(k) = |k|\hat{f}(k)$.

\section{Concluding remarks}

This work establishes a rigorous hydrodynamic theory for premixed flames under Darcy’s law, incorporating non-unity Lewis numbers and conductive heat losses. Using large activation-energy asymptotics for a one-step Arrhenius chemistry model and a systematic multiple-scale framework, we derived the interfacial jump conditions from first principles. These relations introduce finite-flame-thickness corrections to mass flux and pressure continuity that were missing from prior phenomenological formulations. A key finding is that the burning rate in the adiabatic case depends on three distinct Markstein numbers, representing curvature, tangential flow strain, and a gravity-induced strain unique to Darcy flows.

Furthermore, we have demonstrated that non-adiabatic and non-unity Lewis number effects fundamentally alter the flame response. Unlike classical unconfined flames where curvature and tangential strain Markstein numbers coincide for a one-step chemistry model, we show that they are unequal under Darcy’s law $(\mc M_c\neq \mc M_t)$. This decoupling stems from the leading-order tangential velocity discontinuity and gravitational force jump permitted by Darcy’s law, which drastically modify the local flow field in a manner forbidden by Navier--Stokes hydrodynamics. We note that this formulation is based on a one-step Arrhenius mechanism where the flame surface is pinned to the reaction sheet. For complex chemistry model, the flame surface position is not to be identified with the maximum-temperature surface, as discussed in~\cite{clavin2011curved}, which consequently leads to the unequal values for the two Markstein numbers. The origin of inequality under Darcy's law is of different nature.  The relative magnitude of these parameters in our model dictates that $\mc M_t$ is greater than or less than $\mc M_c$ when $Le$ is less than or greater than $Le_*$, where $Le_*$ is greater than unity (typically $Le_* \approx 1.5$). For sub-unity Lewis number mixtures (such as lean hydrogen), the tangential strain contribution thus dominates, making the local burning rate in confined channels uniquely sensitive to tangential straining rather than flame curvature. This highlights that premixed flames subject to pure tangential straining, such as flames in stagnation-point flows, which have been investigated  in our recent work~\cite{rajamanickam2026premixed},  introduce new insights into flame behaviour under Darcy's law.

In summary, the present Darcy-law framework and its explicit Markstein-number formulas offer a precise means of describing flame stability thresholds in porous media and Hele-Shaw cells—an analytical tool previously lacking in the literature. Crucially, it establishes a consistent baseline for future theoretical explorations in  confined combustion. The primary attractiveness of the hydrodynamic model derived herein is that the flow remains potential for any arbitrary value of the density ratio. Since such potential flows are amenable to a vast array of well-established mathematical tools, they offer a fertile ground for further theoretical analysis. In fact, modelling the flame front based on potential flow configurations has attracted significant interest in the past~\cite{Sivashinsky1987,joulin1992tentative,Frankel1990,denet2002potential,Joulin2010PotentialFlow} due to its analytical tractability, though not from a Darcy's-law perspective. Most notably, Frankel~\cite{Frankel1990} derived an invariant evolution equation for arbitrary density ratios under potential flows with the simplified assumption of a constant burning rate $\dot{m}^-=1$. It would be highly interesting to extend that mathematical framework to our Darcy model to incorporate the newly derived Markstein corrections and interfacial jump conditions.

\section*{Acknowledgments}

This work was supported by UK EPSRC through Grant No.~APP39756.

\section*{Appendix: Integral formulas associated with heat-loss terms}
The integrals $\mc J_7$ to $\mc J_{13}$, associated with heat-loss terms in the expressions~\eqref{Mc}-\eqref{N} are given by
\begin{align}
  \mc J_7 &= \mc I_0'''-\mc I_4''+\lambda_f (\mc I_c+\lambda_f), \qquad
 \mc  J_8 = \lambda_f(\mc I_1'''+\mc I_c - \rho_f \lambda_f),\qquad
  \mc J_9 = \lambda_f(\mc I_1'''+\mc I_c - \rho_f \lambda_f) + \mc I_2'''+\lambda_f (\mc I_t-\mc I_c), \\
  \mc J_{10} &= \mc I_3'''+\lambda_f \mc I_g, \qquad
  \mc J_{11} = \mc I_0'''-\mc I_4'' + \mc I_4''' - \lambda_f\mc I_1''', \qquad
  \mc J_{12} = \mc I_5'''+ \frac{\rho_f\lambda_f^2}{\mu_f}(\mu_f-1), \qquad
  \mc J_{13} = \mc I_6'''+ \frac{\rho_f\lambda_f^2}{\mu_f}(1-\rho_f)
\end{align}
where 
\begin{align}
    \mc I_0'''&=\int_0^1   \left\{\rho_0 L + \int_0^\theta \frac{\lambda'}{\theta'}(1-\rho')d\theta'\right\}\lambda d\theta, \qquad \mc I_4''  =\int_0^1   \left\{\int_0^\theta \lambda'd\theta'\right\}\frac{\lambda}{\theta} d\theta, \\
    \mc I_1''' &=\int_0^1   \left\{2\rho L + (2+\ln \theta)\int_0^\theta \frac{\lambda'}{\theta'}(1-\rho')d\theta'\right\}\lambda d\theta, \qquad \mc I_2'''=\int_0^1   \left\{\int_0^\theta \frac{\rho' \lambda'}{\mu'\theta'}(\mu'-1) d\theta' - \frac{\rho}{\mu}(\mu-1)L \right\}\lambda d\theta, \\
    \mc I_3''' & =\int_0^1   \left\{\int_0^\theta \frac{\rho' \lambda'}{\mu'\theta'}(1-\rho') d\theta' - \frac{\rho}{\mu}(1-\rho)L \right\}\lambda d\theta, \qquad 
    \mc I_4''' =  \int_0^1 \left\{\rho L - \lambda\ln\theta + (\ln\theta+1)\int_0^\theta \frac{\lambda'}{\theta'}(1-\rho')d\theta'\right\} \lambda d\theta, \\
    \mc I_5''' &= \int_0^1 \left\{(\ln\theta+1)\int_0^\theta \frac{\rho' \lambda'}{\mu'\theta'}(\mu'-1) d\theta'-\frac{\rho}{\mu}(\mu-1)L\right\}\lambda d\theta, \\ \mc I_6''' &= \int_0^1 \left\{(\ln\theta+1)\int_0^\theta \frac{\rho' \lambda'}{\mu'\theta'}(1-\rho') d\theta'-\frac{\rho}{\mu}(1-\rho)L\right\}\lambda d\theta, \qquad L(\theta) = \int_\theta^1 \frac{\lambda'}{\theta'}d\theta'.
\end{align}


\bibliography{references}

\end{document}